\documentclass{aastex63}
\hypersetup{linkcolor=red,citecolor=purple,filecolor=blue,urlcolor=magenta}

\submitjournal{AJ}

\begin{document}

\shorttitle{Extra-Tidal RR Lyrae in NGC5024 \& NGC5053}
\shortauthors{Ngeow et al.}

\title{A Search for Extra-Tidal RR Lyrae in the Globular Cluster NGC 5024 and NGC 5053}

\correspondingauthor{C.-C. Ngeow}
\email{cngeow@astro.ncu.edu.tw}

\author{Chow-Choong Ngeow}
%\author[0000-0001-8771-7554]{Chow-Choong Ngeow}
\affil{Graduate Institute of Astronomy, National Central University, 300 Jhongda Road, 32001 Jhongli, Taiwan}

\author{Justin Belecki}
\affiliation{Caltech Optical Observatories, California Institute of Technology, Pasadena, CA 91125, USA}

\author{Rick Burruss}
\affiliation{Caltech Optical Observatories, California Institute of Technology, Pasadena, CA 91125, USA}

\author{Andrew J. Drake}
\affiliation{Division of Physics, Mathematics, and Astronomy, California Institute of Technology, Pasadena, CA 91125, USA}

\author{Matthew J. Graham}
%\author[0000-0002-3168-0139]{Matthew J. Graham}
\affiliation{Division of Physics, Mathematics, and Astronomy, California Institute of Technology, Pasadena, CA 91125, USA}

\author{David~L.\ Kaplan}
%\author[0000-0001-6295-2881]{David~L.\ Kaplan}
\affiliation{Center for Gravitation, Cosmology and Astrophysics, Department of Physics, University of Wisconsin--Milwaukee, Milwaukee, WI 53201, USA}

\author{Thomas Kupfer}
%\author[0000-0002-6540-1484]{Thomas Kupfer}
\affiliation{Kavli Institute for Theoretical Physics, University of California, Santa Barbara, CA 93106, USA}

\author{Ashish Mahabal}
%\author[0000-0003-2242-0244]{Ashish Mahabal}
\affiliation{Division of Physics, Mathematics, and Astronomy, California Institute of Technology, Pasadena, CA 91125, USA}
\affiliation{Center for Data Driven Discovery, California Institute of Technology, Pasadena, CA 91125, USA}

\author{Frank J. Masci}
%\author[0000-0002-8532-9395]{Frank J. Masci}
\affiliation{IPAC, California Institute of Technology, 1200 E. California Blvd, Pasadena, CA 91125, USA}

\author{Reed Riddle}
\affiliation{Caltech Optical Observatories, California Institute of Technology, Pasadena, CA 91125, USA}

\author{Hector Rodriguez}
\affiliation{Caltech Optical Observatories, California Institute of Technology, Pasadena, CA 91125, USA}

\author{Ben Rusholme}
%\author[0000-0001-7648-4142]{Ben Rusholme}
\affiliation{IPAC, California Institute of Technology, 1200 E. California Blvd, Pasadena, CA 91125, USA}

\begin{abstract}

  Recently, \citet{kundu2019} reported that the globular cluster
  NGC~5024 (M53) possesses five extra-tidal RR Lyrae. In fact, four of
  them were instead known members of a nearby globular cluster
  NGC~5053. The status of the remaining extra-tidal RR Lyrae is
  controversial depending on the adopted tidal radius of NGC~5024. We
  have also searched for additional  extra-tidal RR Lyrae within an
  area of $\sim8$~deg$^2$ covering both globular clusters. This
  includes other known RR Lyrae within the search area, as well as
  stars that fall within the expected range of magnitudes and colors
  for RR Lyrae (and yet outside the cutoff of $2/3$ of the tidal radii
  of each globular clusters for something to be called ``extra-tidal'') if they were extra-tidal RR Lyrae candidates for NGC~5024 or NGC~5053. Based on the the proper motion information and their locations on the color-magnitude diagram, none of the known RR Lyrae  belong to the extra-tidal RR Lyrae of either globular clusters. In the cases where the stars satisfied the magnitude and color ranges of RR Lyrae, analysis of time series data taken from the Zwicky Transient Facility do not reveal periodicities, suggesting that none of these stars are RR Lyrae. We conclude that there are no extra-tidal RR Lyrae associated with either NGC~5024 or NGC~5053 located within our search area.

\end{abstract}

\section{Introduction}

As an ancient population that is orbiting around our Galaxy in its halo, globular clusters could leave tidal tails along their orbits. Indeed, tidal tails have already been detected in some globular clusters such as Palomar 5 \citep{odenkirchen2001,rockosi2002}, NGC~5466 \citep{belokurov2006}, Palomar 1 \citep{no2010}, Palomar 14 \citep{sollima2011}, Palomar 15 \citep{myeong2017}, Eridanus \citep{myeong2017}, NGC~7492 \citep{navarrete2017} and M5 \citep{grillmair2019}. The existence of tidal tails from globular clusters were also supported from theoretical modeling and simulations \citep[for examples, see][]{combes1999,yim2002,dehnen2004,cd2005,lee2006,fellhauer2007,montouri2007,hozumi2015}. The tidally stripped stars in the tails are preferentially low mass stars \citep{combes1999,baumgardt2003}, which could include RR Lyrae \citep{jordi2010}. As low-mass high-amplitude pulsating stars, RR Lyrae can be easily identified from time domain surveys based on their characteristic light curve shapes. RR Lyrae are precise distance indicators hence distances to the tidal tails can be constrained if the extra-tidal RR Lyrae can be found in the associated tidal tails. Furthermore, RR Lyrae are more luminous than the main-sequence stars at similar mass, hence they can be used to reveal the presence of tidal tails around distant globular clusters. In this regards, it is important and interesting to search for extra-tidal RR Lyrae associated with  globular clusters. As demonstrated in \citet{kunder2018} and \citet{minniti2018}, finding of extra-tidal RR Lyrae could be used to constrain history of orbital motion and dynamic processes of the parent globular clusters. 

\begin{figure*}
  \plotone{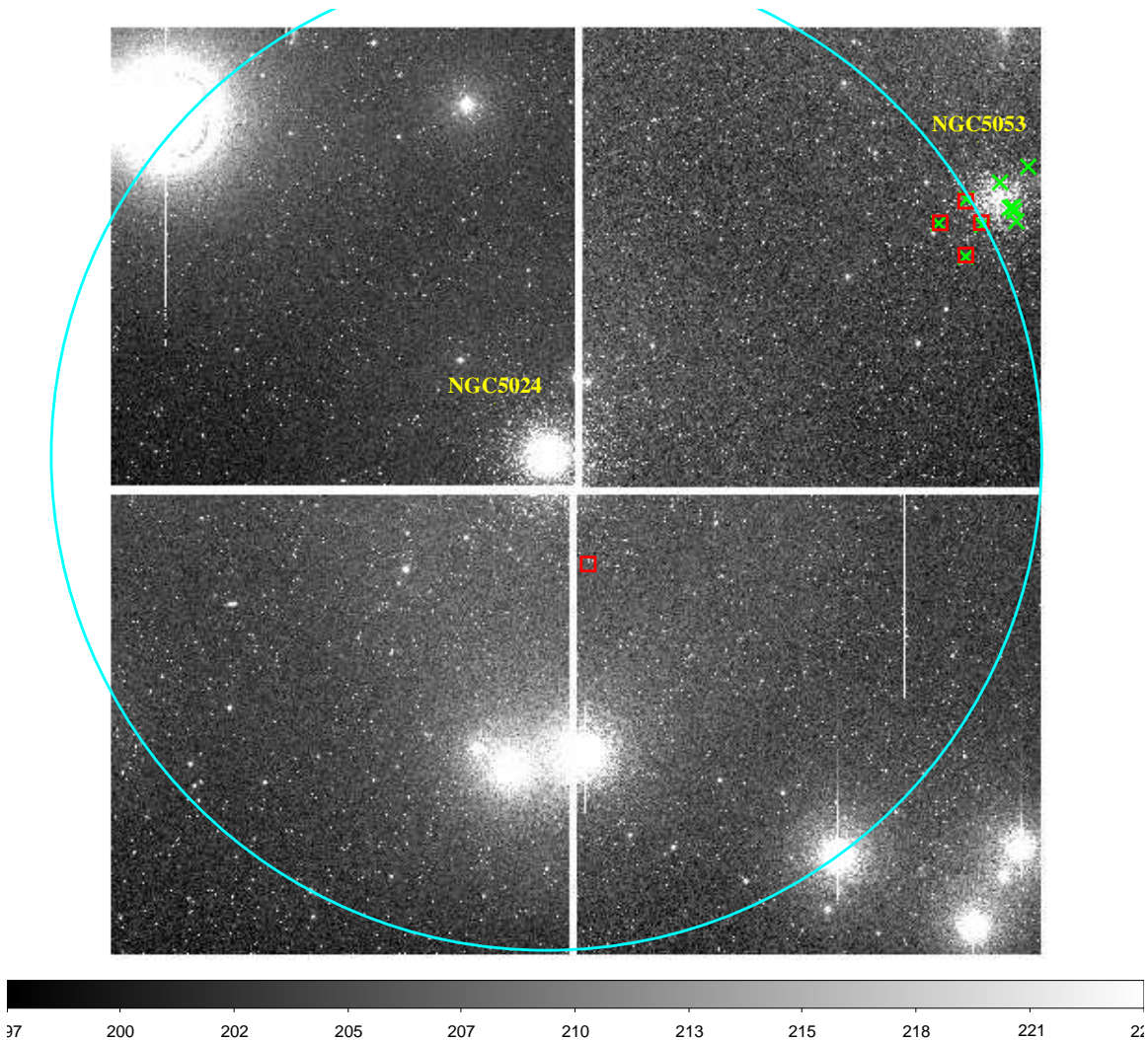}
  \caption{ZTF mosaic  (see Section 3 for more details on ZTF)
    $r$-band reference images, with ZTFID of 1573 and CCD 15, that
    display the positions of NGC~5024 and NGC~5053. The cyan circle
    indicates the radius of $3\times r_t$ of NGC~5024 as adopted in
    \citet{kundu2019}. The red squares represent the extra-tidal RR
    Lyrae identified in \citet{kundu2019}, while the green crosses are
    known RR Lyrae of NGC~5053 taken from Clement's catalog \citep{clement2001,clement2017}.}\label{fig1}
\end{figure*}

Extra-tidal RR Lyrae stars have been found around several globular
clusters. Based on wide field imaging, \citet{ft2015} reported about a
dozen RR Lyrae located at a distance similar to $\omega$~Centauri
(NGC~5139), which are also located outside the tidal radius ($r_t$) of
this cluster. However the authors suggested they are unlikely to be
associated with the tidal debris of $\omega$~Centauri. In contrast,
eight extra-tidal RR Lyrae candidates were found for NGC~6441, as
their radial velocities are consistent with the cluster and they are
located between 1 to 3 tidal radii from the cluster
\citep{kunder2018}. Using Gaia DR2 \citep[Data Release
  2;][]{gaia2016,gaia2018} data, \citet{minniti2018} discovered that
there is an excess of RR Lyrae outside the tidal radius of M62
(NGC~6266),  which the authors interpreted as tidally stripped RR
Lyrae while the cluster is crossing the Galactic bulge. Recently,
\citet{pw2019} found 17 RR Lyrae that most likely belong to the
stellar stream of Palomar 5, with a few of them previously identified  as potential members  \citep{vivas2001,vivas2006,wu2005}. An independent search of RR Lyrae in the stellar stream of Palomar 5 was also performed in \citet{ibata2017}.

In addition to the individual globular clusters, \citet{kundu2019}
performed a systematic search for extra-tidal RR Lyrae around globular
clusters with Gaia DR2 data and the Gaia DR2 RR Lyrae Catalog
\citep{clementini2019}. Out of the 56 globular clusters, the authors
identify 11 globular clusters possessing extra-tidal RR Lyrae based on
their positions (out to three times the tidal radius of each
clusters), proper motions and  locations in the color-magnitude
diagrams. Among these 11 globular clusters, six globular clusters have
one extra-tidal RR Lyrae while another three clusters have two
(including Palomar 5). The remaining two globular clusters, NGC~5024
and NGC~3201, were found to have 5 and 13 extra-tidal RR Lyrae,
respectively. The extra-tidal RR Lyrae in NGC~5024 merit further
discussion, as four of them are located on one side of the cluster at
a (projected) distance near 3 times the tidal radius (that is,
$55.1$\,arc-minute). This  is close to a nearby globular cluster:
NGC~5053 located at a projected distance of $\sim57.7$\,arc-minute away
from NGC~5024. In fact, these four extra-tidal RR Lyrae of NGC~5024
are instead known RR Lyrae from NGC~5053 as listed in the ``Updated Catalog of Variable Stars in Globular Clusters'' \citep[][hereafter Clement's catalog]{clement2001,clement2017}, as illustrated in Figure \ref{fig1}. Hence the number of extra-tidal RR Lyrae in NGC~5024 should be reduced to one.

Since NGC~5053 is not included in the list of 56 globular clusters
given in \citet{kundu2019}, it may contain additional (uncatalogued)
extra-tidal RR Lyrae. Separately, we wish to ascertain whether  there
could be more extra-tidal RR Lyrae from NGC~5024, in addition to the
one that was identified and discussed above. The goal of this work is to search for (additional) extra-tidal RR Lyrae in NGC~5024 and NGC~5053, given the close proximity of these two globular clusters in the sky. Note that based on imaging observations using the Canada-Hawaii-France Telescope (CFHT), \citet{chun2010} reported there was a tidal bridge feature between NGC~5024 and NGC~5053. However, such a tidal bridge feature was not confirmed in \citet{jordi2010}. Nevertheless, \citet{jordi2010} found extra-tidal halos in both globular clusters, and confirmed the detection of tidal tail for NGC~5053 reported in \citet{lauchner2006}. On the other hand, a possible tidal tail could be present for NGC~5024 \citep{beccari2008} but no conclusive result can be determined..

In Section 2, we compile a list of known and candidate RR Lyrae
collected from the literature, and determine if any of them are extra-tidal RR Lyrae. In Section 3, we search for potential new extra-tidal RR Lyrae using Gaia DR2 data and time series data taken from the Zwicky Transient Facility (ZTF). The conclusion of this work is  presented in Section 4.

\section{Known and Candidate RR Lyrae in the Vicinity of the Clusters}

The tidal radii $r_t$ of NGC~5024 and NGC~5053 were adopted from
\citet{deboer2019} as $22.8\pm1.4$\,arc-minute and
$15.2\pm3.3$\,arc-minute, respectively. For consistency, these tidal
radii were based on the fitted SPES (spherical potential escapers
stitched) model\footnote{In case of NGC~5053, even though the Wilson
  model is a better fit to its number density \citep{deboer2019}, the
  derived tidal radius from Wilson model, $18.1\pm1.5$\,arc-minute, is
  in good agreement with the value based on SPES model.} to the number
density profile constructed from Gaia DR2 and literature data,
converted from parsec to arc-minute using the distances provided in
\citet[][the 2010 edition; hereafter Harris Catalog]{harris1996,
  harris2010}\footnote{We thank T. de Boer for verifying this.}, where
the distancse to NGC~5024 and NGC~5053 are 17.9\,kpc and 17.4\,kpc,
respectively. Because the areas enclosed by the $3r_t$ regions for these two clusters overlapped, we defined a circle with radius of $1.6$\,degree, centered at $(\alpha,\delta)_{J2000}=(198.48568, +17.90798)$\,degree (see Figure \ref{fig_grouB_img}), to search and identify RR Lyrae from the literature within this $\sim8$\,deg$^2$ area.

We collected known and candidate RR Lyrae located within this circle from various catalogs. These catalogs include  Clement's catalog (NGC~5024: 64 stars; NGC~5053: 10 stars), the catalog from the AAVSO International Variable Star Index \citep[VSX;][55 stars]{watson2006}, and from \citet[][91 stars]{sesar2017}. The primary recent sources for RR Lyrae compiled in  Clement's catalogs are: \citet{af2011} for NGC~5024 and \citet{nemec2004} for NGC~5053, respectively. We selected all entries in \citet{sesar2017}  within the pre-defined circle regardless of the final classification scores {\tt S3ab} and {\tt S3c}, hence some of them with low scores were considered as RR Lyrae candidates.\footnote{The scores have a value between 0 and 1, where 1 represents the star has a very high probability being a RR Lyrae, and 0 if the star is not classified as a RR Lyrae. These scores have an associated level of purity and completeness as described in \citet{sesar2017}. For example, {\tt S3ab > 0.8} implies the underlying star is an ab-type RR Lyrae with purity of 91\% and completeness of 77\% at $\sim80$\,kpc.} All of the above were queried via the {\tt SIMBAD's VizieR} service. We have also searched the Gaia DR2 RR Lyrae Catalog \citep[][the {\tt gaiadr2.vary\_rrlyrae} Table; 79 stars]{clementini2019} and RR Lyrae in the Gaia DR2 high-amplitude pulsating stars Catalog \citep[][the {\tt gaiadr2.vari\_classifier\_result} Table; 70 stars]{rimoldini2019} via the ADQL interface from the Gaia archive\footnote{{\tt http://gea.esac.esa.int/archive/}}. We combined the query results from these two tables, using the {\tt source\_id}, for a total of 80 RR Lyrae from Gaia DR2 (hereafter GaiaDR2RRL catalog). Finally, we merged all of the above catalogs, using positional matching, to create a master catalog that contains 125 RR Lyrae and candidates within the  circle mentioned earlier.

\begin{figure*}
  \epsscale{1.0}
  \plotone{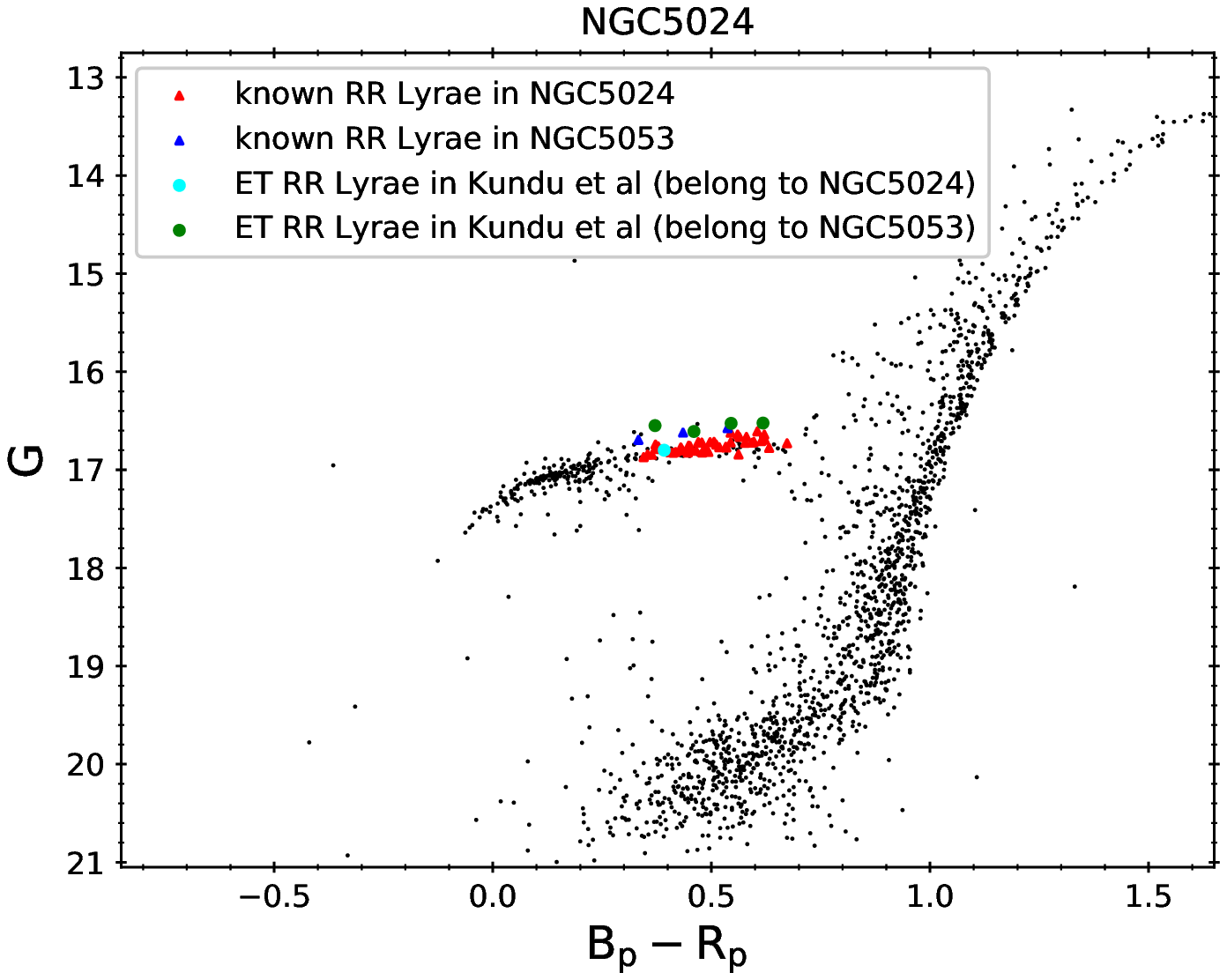}
  \plotone{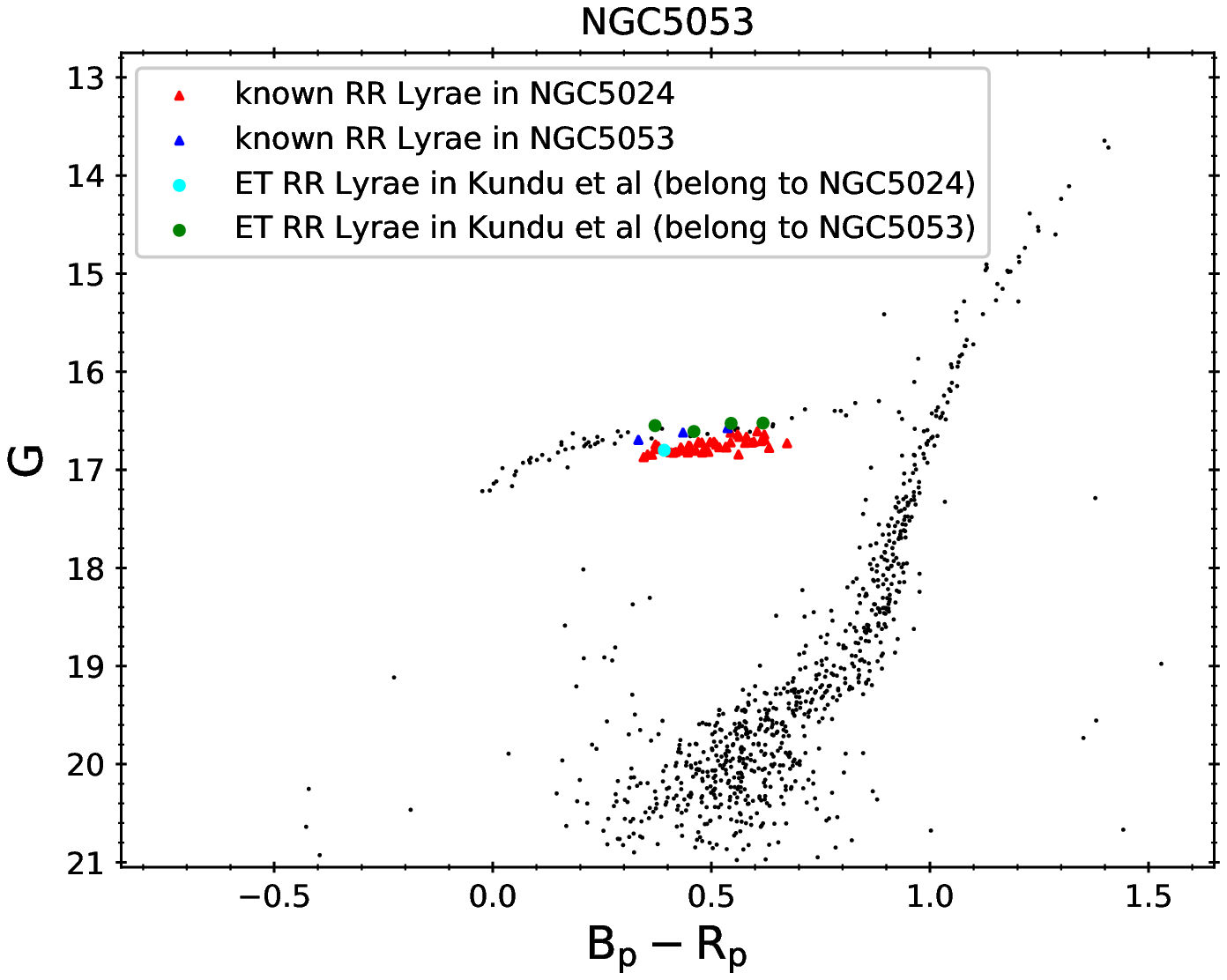}
  \caption{The ``clean'' color-magnitude diagram (CMD) for NGC~5024 (left panel) and NGC~5053 (right panel) based on Gaia photometry. The color-coded symbols represent the known RR Lyrae, based on  Clement's catalog, in these two globular clusters. Extra-tidal RR Lyrae identified in \citet{kundu2019} are abbreviated as ET RR Lyrae (four of them belong to NGC~5053 and are shown as large green points). Procedures for obtaining the ``clean'' CMD were explained in Appendix A. The CMD for NGC~5053 is similar to those presented in \citet{sm1995}.}\label{fig_groupA_cmd}
\end{figure*}

\begin{figure*}
  \epsscale{1.2}
  \plotone{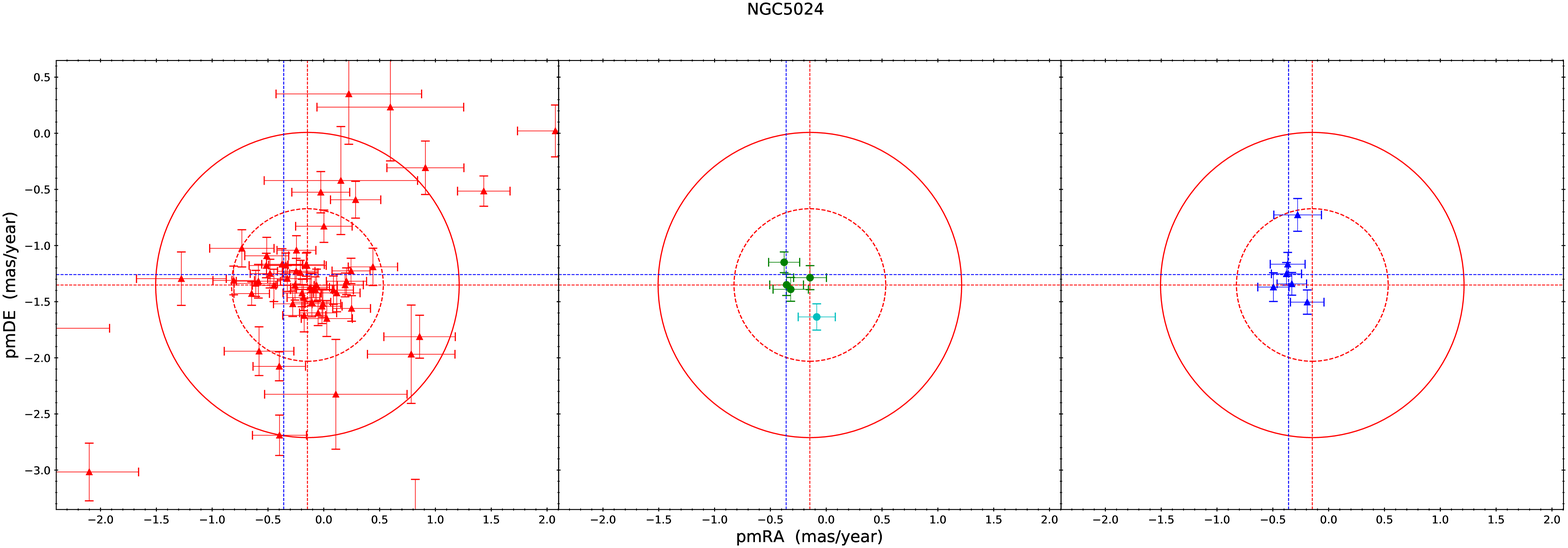}
  \plotone{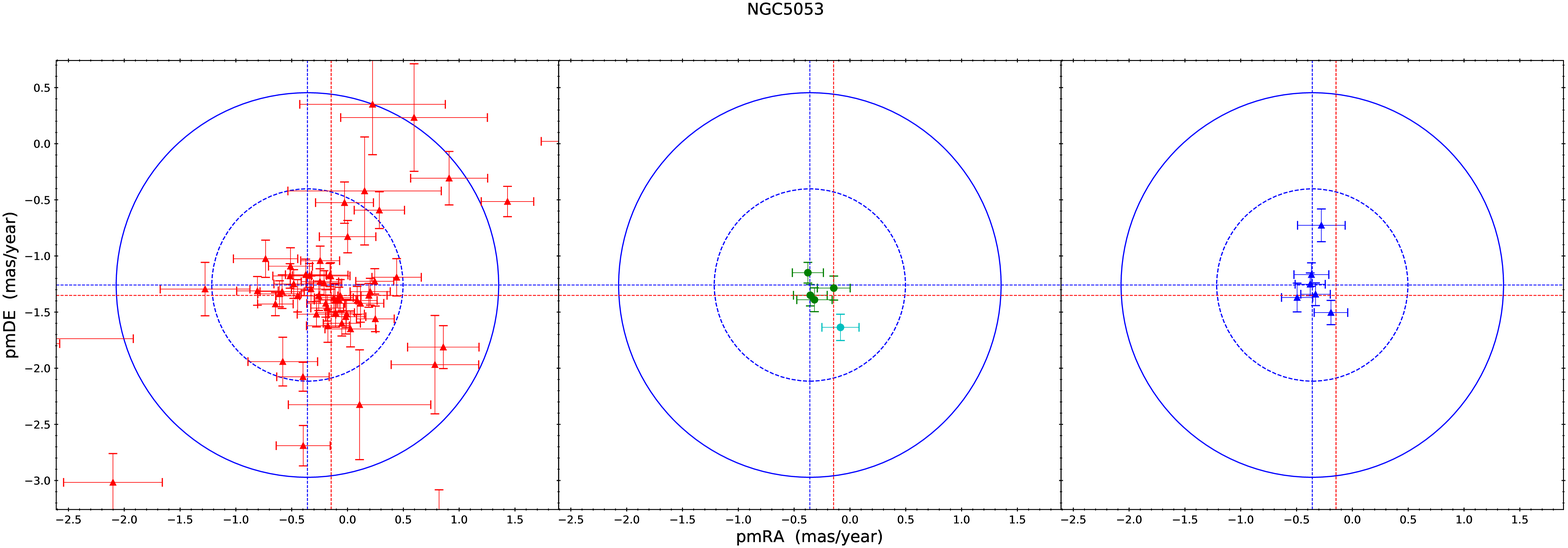}
  \caption{Comparisons of the proper motions of the known RR Lyrae in
    NGC~5024 and NGC~5053 to the values of the two globular clusters themselves. The red and blue dashed lines in each panels represent the measured proper motions of NGC~5024 and NGC~5053, respectively, adopted from \citet{gaia2018gc}. The dashed and solid circles represent the selection criteria of $\Delta_{\mathrm{PM}} = \mathrm{pmT}/2$ and $\Delta_{\mathrm{PM}} = \mathrm{pmT}$, respectively; see Appendix A for the definitions of $\Delta_{\mathrm{PM}}$ and $\mathrm{pmT}$. The colored symbols are same as in Figure \ref{fig_groupA_cmd}. For better visualization, we divided the known RR Lyrae to those belong to NGC~5024 (left panels; red points), NGC~5053 (right panels; blue points), and the extra-tidal RR Lyrae (middle panels; green and cyan points).}\label{fig_groupA_pm}
\end{figure*}

A further examination of these 125 RR Lyrae revealed that  29 RR Lyrae candidates do not have any counterparts in either  Clement's catalog, the VSX catalog or the GaiaDR2RRL catalog. All of them have a classification scores {\tt S3ab} and {\tt S3c} smaller than 0.52 in the \citet{sesar2017} catalog, with brightness fainter than $\sim18.5$\,mag in the $r$-band or Gaia's $G$-band. We believe they are either mis-classified or background RR Lyrae that are unrelated to NGC~5024 and NGC~5053 (the RR Lyrae in these two globular clusters should be brighter than $\sim17.5$\,mag in $r$- or $G$-band), hence we removed them from our master catalog (a detailed investigation of them is beyond the scope of this work). The remainder of the 96 RR Lyrae in our master catalog will be divided into two groups, and analyzed with criteria similar to those outlined in \citet{kundu2019}, as presented in the following subsections.

\subsection{Group A: Known Members in NGC~5024 and NGC~5053}

%\startlongtable
\begin{deluxetable*}{lccccccccc}
  %\movetableright=-1in
  %\tabletypesize{\scriptsize}
  \tabletypesize{\tiny}
  \tablecaption{Known RR Lyrae in NGC~5024 and NGC~5053 (Group A).\label{tab1}}
  \tablewidth{0pt}
  \tablehead{
    \colhead{Name\tablenotemark{a}} &
    \colhead{$\alpha_{J2000}$} &
    \colhead{$\delta_{J2000}$} &
    \colhead{$\Delta_{5024}$\tablenotemark{b}} &
    \colhead{$\Delta_{5053}$\tablenotemark{b}} &
    \colhead{Gaia DR2 ID} &
    \colhead{$G$\tablenotemark{c}} &
    \colhead{$B_p-R_p$\tablenotemark{c}} &
    \colhead{pmRA\tablenotemark{d}} &
    \colhead{pmDE\tablenotemark{d}} }
  \startdata
  NGC5024 V61 & 198.22967 & 18.17014 & 0.12 & 57.76 & 3938022017256098432 & -99.99 & -99.99 & $0.224\pm0.652$ & $0.351\pm0.449$ \\
  NGC5024 V57 & 198.23150 & 18.16617 & 0.14 & 57.56 & 3938022017253298816 & 16.620 & -99.99 & $0.820\pm0.549$ & $-3.508\pm0.425$ \\
  NGC5024 V63 & 198.23454 & 18.16686 & 0.26 & 57.43 & 3937271394410272128 & -99.99 & -99.99 & $0.596\pm0.657$ & $0.233\pm0.479$ \\
  NGC5024 V72 & 198.23308 & 18.16453 & 0.27 & 57.43 & 3937271394411045760 & -99.99 & -99.99 & $2.074\pm0.339$ & $0.021\pm0.231$ \\
  NGC5024 V71 & 198.22658 & 18.16500 & 0.28 & 57.77 & 3938022017255714688 & -99.99 & -99.99 & $-2.708\pm0.789$ & $-1.737\pm0.813$ \\
  NGC5024 V91 & 198.22342 & 18.17044 & 0.41 & 58.08 & 3938022017256046592 & -99.99 & -99.99 & $2.884\pm0.445$ & $1.536\pm0.333$ \\
  NGC5024 V53 & 198.23263 & 18.17544 & 0.46 & 57.77 & 3938022395210336128 & -99.99 & -99.99 & $-99.99$ & $-99.99$ \\
  NGC5024 V54 & 198.22629 & 18.17542 & 0.49 & 58.09 & 3938022017256105344 & 16.633 & -99.99 & $-3.039\pm0.587$ & $5.163\pm0.476$ \\
  NGC5024 V62 & 198.22500 & 18.17494 & 0.50 & 58.14 & 3938022017255910528 & -99.99 & -99.99 & $0.153\pm0.687$ & $-0.421\pm0.481$ \\
  NGC5024 V46 & 198.22721 & 18.17694 & 0.55 & 58.09 & 3938022017255848704 & 16.622 & 0.544 & $-0.734\pm0.288$ & $-1.025\pm0.166$ \\
  NGC5024 V52 & 198.23300 & 18.17697 & 0.55 & 57.80 & 3938022395212860928 & 16.782 & -99.99 & $0.783\pm0.392$ & $-1.968\pm0.438$ \\
  NGC5024 V58 & 198.23167 & 18.15861 & 0.58 & 57.33 & 3937271394411059072 & 16.665 & 0.564 & $0.910\pm0.345$ & $-0.307\pm0.238$ \\
  NGC5024 V45 & 198.22983 & 18.15761 & 0.63 & 57.39 & 3937271394410013312 & 16.640 & 0.560 & $-0.027\pm0.259$ & $-0.525\pm0.184$ \\
  NGC5024 V60 & 198.23742 & 18.16014 & 0.63 & 57.09 & 3937271394411025920 & 16.385 & -99.99 & $-99.99$ & $-99.99$ \\
  NGC5024 V64 & 198.21883 & 18.17014 & 0.66 & 58.31 & 3938022017255783424 & -99.99 & -99.99 & $-2.644\pm1.265$ & $1.637\pm0.863$ \\
  NGC5024 V55 & 198.22275 & 18.17683 & 0.67 & 58.30 & 3938022085972624768 & 16.484 & -99.99 & $0.001\pm0.253$ & $-0.828\pm0.144$ \\
  NGC5024 V56 & 198.22371 & 18.15722 & 0.75 & 57.69 & 3937271360050576512 & 16.778 & -99.99 & $-2.101\pm0.442$ & $-3.017\pm0.257$ \\
  NGC5024 V92 & 198.22879 & 18.18064 & 0.75 & 58.12 & 3938022395213189248 & -99.99 & -99.99 & $-1.276\pm0.402$ & $-1.295\pm0.238$ \\
  NGC5024 V59 & 198.23608 & 18.15578 & 0.82 & 57.03 & 3937271394411208192 & -99.99 & -99.99 & $0.108\pm0.638$ & $-2.325\pm0.489$ \\
  NGC5024 V44 & 198.21525 & 18.16683 & 0.86 & 58.39 & 3938021982896346240 & 16.718 & 0.470 & $-0.399\pm0.235$ & $-2.076\pm0.129$ \\
  NGC5024 V51 & 198.23996 & 18.18047 & 0.92 & 57.56 & 3938022395210481408 & -99.99 & -99.99 & $1.433\pm0.235$ & $-0.515\pm0.135$ \\
  NGC5024 V43 & 198.22117 & 18.18208 & 0.98 & 58.54 & 3938022085975189504 & 16.664 & 0.580 & $0.285\pm0.225$ & $-0.592\pm0.164$ \\
  NGC5024 V31 & 198.24821 & 18.16803 & 1.03 & 56.78 & 3937271772365640960 & 16.608 & 0.605 & $-0.807\pm0.186$ & $-1.311\pm0.128$ \\
  NGC5024 V41 & 198.23646 & 18.18569 & 1.11 & 57.89 & 3938022395210877312 & 16.760 & 0.378 & $0.203\pm0.180$ & $-1.318\pm0.119$ \\
  NGC5024 V42 & 198.21058 & 18.17225 & 1.15 & 58.78 & 3938022051615475712 & 16.640 & 0.621 & $-2.918\pm0.340$ & $-1.781\pm0.190$ \\
  NGC5024 V37 & 198.21783 & 18.18483 & 1.22 & 58.78 & 3938022085972862848 & 16.693 & 0.622 & $-0.247\pm0.168$ & $-1.229\pm0.115$ \\
  NGC5024 V09 & 198.25029 & 18.15697 & 1.33 & 56.35 & 3937271321396679936 & 16.777 & 0.632 & $-0.447\pm0.193$ & $-1.355\pm0.144$ \\
  NGC5024 V18 & 198.20250 & 18.17033 & 1.58 & 59.13 & 3938022047320224768 & 16.822 & 0.406 & $-0.580\pm0.312$ & $-1.941\pm0.217$ \\
  NGC5024 V08 & 198.25171 & 18.18475 & 1.58 & 57.11 & 3938022429571836288 & 16.772 & 0.534 & $0.858\pm0.320$ & $-1.812\pm0.191$ \\
  NGC5024 V40 & 198.23276 & 18.19852 & 1.83 & 58.46 & 3938022498291497088 & 16.843 & 0.354 & $-0.051\pm0.166$ & $-1.600\pm0.113$ \\
  NGC5024 V07 & 198.25363 & 18.19165 & 1.94 & 57.23 & 3938022429572023424 & 16.844 & 0.562 & $0.440\pm0.221$ & $-1.190\pm0.166$ \\
  NGC5024 V24 & 198.19700 & 18.15900 & 1.97 & 59.09 & 3938021948535632768 & -99.99 & -99.99 & $-0.077\pm0.168$ & $-1.389\pm0.110$ \\
  NGC5024 V06 & 198.26630 & 18.17220 & 2.07 & 56.01 & 3937271733713546624 & 16.706 & 0.614 & $0.027\pm0.229$ & $-1.650\pm0.160$ \\
  NGC5024 V25 & 198.26841 & 18.17703 & 2.24 & 56.05 & 3937271733710749056 & 16.720 & 0.544 & $-0.176\pm0.192$ & $-1.622\pm0.147$ \\
  NGC5024 V23 & 198.25978 & 18.14332 & 2.25 & 55.49 & 3937271119532519808 & 16.772 & 0.430 & $-0.397\pm0.242$ & $-2.690\pm0.180$ \\
  NGC5024 V32 & 198.19893 & 18.14327 & 2.33 & 58.56 & 3938021841161792640 & 16.758 & 0.451 & $-0.008\pm0.159$ & $-1.517\pm0.121$ \\
  NGC5024 V10 & 198.19053 & 18.18207 & 2.41 & 60.06 & 3938022253478651392 & 16.722 & 0.478 & $-0.327\pm0.175$ & $-1.294\pm0.115$ \\
  NGC5024 V38 & 198.23812 & 18.12791 & 2.46 & 56.15 & 3937270977799285376 & 16.714 & 0.597 & $-0.646\pm0.159$ & $-1.428\pm0.104$ \\
  NGC5024 V03 & 198.21410 & 18.12926 & 2.51 & 57.41 & 3937271149597965312 & 16.818 & 0.493 & $-0.341\pm0.153$ & $-1.174\pm0.107$ \\
  NGC5024 V29 & 198.26779 & 18.14637 & 2.51 & 55.17 & 3937271497489647616 & -99.99 & -99.99 & $-0.482\pm0.177$ & $-1.250\pm0.128$ \\
  NGC5024 V11 & 198.18912 & 18.15054 & 2.57 & 59.25 & 3938021909878970752 & 16.765 & 0.533 & $-0.151\pm0.173$ & $-1.178\pm0.118$ \\
  NGC5024 V47 & 198.21008 & 18.20686 & 2.59 & 59.83 & 3938022738809664000 & 16.742 & 0.372 & $-0.245\pm0.174$ & $-1.042\pm0.130$ \\
  NGC5024 V33 & 198.18278 & 18.17024 & 2.71 & 60.12 & 3938022154692482944 & 16.729 & 0.673 & $0.086\pm0.179$ & $-1.397\pm0.126$ \\
  NGC5024 V01 & 198.23478 & 18.12049 & 2.87 & 56.12 & 3937270982093531648 & 16.770 & 0.518 & $0.249\pm0.169$ & $-1.560\pm0.114$ \\
  NGC5024 V19 & 198.27926 & 18.15730 & 2.87 & 54.92 & 3937271531849398400 & 16.802 & 0.434 & $-0.111\pm0.187$ & $-1.393\pm0.143$ \\
  NGC5024 V35 & 198.26011 & 18.21047 & 3.06 & 57.50 & 3938022567010985984 & 16.824 & 0.446 & $-0.374\pm0.182$ & $-1.167\pm0.139$ \\
  NGC5024 V02 & 198.20952 & 18.11687 & 3.30 & 57.31 & 3937269676423468672 & 16.788 & 0.372 & $-0.609\pm0.194$ & $-1.337\pm0.116$ \\
  NGC5024 V04 & 198.18288 & 18.12395 & 3.78 & 58.87 & 3938020367987656320 & 16.800 & 0.432 & $0.191\pm0.163$ & $-1.351\pm0.106$ \\
  NGC5024 V17 & 198.16807 & 18.19833 & 3.98 & 61.66 & 3938023082407017088 & 16.806 & 0.465 & $0.244\pm0.170$ & $-1.226\pm0.113$ \\
  NGC5024 V16 & 198.19250 & 18.11085 & 4.06 & 58.04 & 3937269642063727360 & 16.871 & 0.345 & $-0.195\pm0.170$ & $-1.423\pm0.122$ \\
  NGC5024 V27 & 198.17261 & 18.12322 & 4.25 & 59.38 & 3938020333627916928 & 16.723 & 0.593 & $-0.515\pm0.154$ & $-1.178\pm0.109$ \\
  NGC5024 V34 & 198.19045 & 18.10722 & 4.30 & 58.05 & 3937269642063726080 & -99.99 & -99.99 & $-0.123\pm0.174$ & $-1.372\pm0.125$ \\
  NGC5024 V36 & 198.26373 & 18.25287 & 5.43 & 58.72 & 3939524087576094848 & 16.825 & 0.413 & $-0.065\pm0.186$ & $-1.354\pm0.144$ \\
  NGC5024 V15 & 198.30162 & 18.23205 & 5.59 & 56.23 & 3939523915777402368 & 16.847 & 0.364 & $-0.511\pm0.197$ & $-1.091\pm0.164$ \\
  NGC5024 V05 & 198.16285 & 18.09508 & 5.83 & 59.19 & 3937269573344246528 & 16.725 & 0.577 & $-0.158\pm0.162$ & $-1.170\pm0.103$ \\
  NGC5024 V26 & 198.14894 & 18.08894 & 6.64 & 59.78 & 3937269191089937536 & 16.753 & 0.448 & $-0.278\pm0.173$ & $-1.519\pm0.113$ \\
  NGC5024 V20 & 198.28781 & 18.07121 & 6.68 & 52.10 & 3937269809565198720 & 16.801 & 0.449 & $-0.107\pm0.169$ & $-1.505\pm0.130$ \\
  NGC5024 V14 & 198.33562 & 18.11195 & 6.89 & 50.79 & 3937270359321073792 & 16.784 & 0.488 & $-0.016\pm0.180$ & $-1.541\pm0.154$ \\
  NGC5024 V28 & 198.17547 & 18.27716 & 7.25 & 63.72 & 3938026930697675648 & 16.715 & 0.506 & $-0.210\pm0.158$ & $-1.241\pm0.110$ \\
  NGC5024 V21 & 198.35868 & 18.16208 & 7.33 & 51.20 & 3938772017327358336 & 16.818 & 0.418 & $-0.586\pm0.198$ & $-1.318\pm0.150$ \\
  NGC5024 V12 & 198.34930 & 18.22126 & 7.50 & 53.63 & 3938773048119562496 & 16.742 & 0.511 & $-0.258\pm0.171$ & $-1.355\pm0.131$ \\
  NGC5024 V30 & 198.25068 & 18.03440 & 8.11 & 53.16 & 3937268160297720576 & 16.824 & 0.479 & $-0.182\pm0.147$ & $-1.464\pm0.116$ \\
  NGC5024 V13 & 198.36767 & 18.08693 & 9.23 & 48.46 & 3937267133802945024 & 16.718 & 0.497 & $0.116\pm0.209$ & $-1.422\pm0.168$ \\
  NGC5024 V48 & 198.30983 & 18.36777 & 12.81 & 60.85 & 3939527386110983424\tablenotemark{e} & 16.800 & 0.392 & $-0.085\pm0.166$ & $-1.636\pm0.117$ \\
  NGC5053 V10 & 199.13554 & 17.71290 & 58.45 & 1.50 & 3938494459361028992 & -99.99 & -99.99 & $-0.278\pm0.213$ & $-0.727\pm0.146$ \\
  NGC5053 V08 & 199.14179 & 17.71116 & 58.82 & 1.78 & 3938494463656186880 & 16.620 & 0.435 & $-0.367\pm0.156$ & $-1.166\pm0.105$ \\
  NGC5053 V04 & 199.11674 & 17.66565 & 58.91 & 2.09 & 3938493948260097024 & 16.501 & -99.99 & $-0.495\pm0.140$ & $-1.370\pm0.128$ \\
  NGC5053 V06 & 199.14425 & 17.71869 & 58.73 & 2.11 & 3938494566735402880 & 16.695 & 0.333 & $-0.192\pm0.150$ & $-1.504\pm0.108$ \\
  NGC5053 V03 & 199.14842 & 17.73773 & 58.43 & 3.03 & 3938682273986049792 & 16.575 & 0.537 & $-0.378\pm0.135$ & $-1.252\pm0.101$ \\
  NGC5053 V07 & 199.08220 & 17.74399 & 54.89 & 3.16 & 3938682445784738688\tablenotemark{e} & 16.549 & 0.371 & $-0.318\pm0.158$ & $-1.390\pm0.106$ \\
  NGC5053 V02 & 199.05154 & 17.69631 & 54.77 & 3.51 & 3938681827309432448\tablenotemark{e} & 16.609 & 0.460 & $-0.145\pm0.147$ & $-1.286\pm0.107$ \\
  NGC5053 V05 & 199.17174 & 17.63628 & 62.52 & 5.10 & 3938492883108232960 & 16.513 & 0.617 & $-0.330\pm0.132$ & $-1.341\pm0.101$ \\
  NGC5053 V01 & 198.99724 & 17.74100 & 50.73 & 7.05 & 3938682995540541440\tablenotemark{e} & 16.526 & 0.545 & $-0.355\pm0.151$ & $-1.350\pm0.095$ \\
  NGC5053 V09 & 199.04957 & 17.80299 & 51.64 & 7.15 & 3938686637672823424\tablenotemark{e} & 16.523 & 0.618 & $-0.377\pm0.139$ & $-1.149\pm0.092$ 
  \enddata
  \tablenotetext{a}{Known RR Lyrae adopted from Clement's Catalog. The number $-99.99$ means no data.}
  \tablenotetext{b}{$\Delta$ represents the angular distance, in arc-minutes, for a given RR Lyrae to the center of the cluster.}
  \tablenotetext{c}{Intensity mean magnitudes and colors taken from the Gaia DR2 RR Lyrae Catalog (i.e. the {\tt gaiadr2.vary\_rrlyrae} Table).}
  \tablenotetext{d}{Proper motions in mas~yr$^{-1}$ taken from Gaia DR2 main catalog.}
  \tablenotetext{e}{extra-tidal RR Lyrae for NGC~5024 identified in \citet{kundu2019}.}
\end{deluxetable*}

In Table \ref{tab1} we summarize the basic information for the 74 known RR Lyrae located in NGC~5024 and NGC~5053. Positions of these RR Lyrae in the color-magnitude diagrams (CMD) are presented in Figure \ref{fig_groupA_cmd}. The stars, including the extra-tidal RR Lyrae identified in \citet{kundu2019}, are undoubtedly RR Lyrae stars as they are located as expected on horizontal branch on the CMD's of these two globular clusters. The four extra-tidal RR Lyrae that belong to NGC~5053 (green points in Figure \ref{fig_groupA_cmd}), but mis-identified as members of NGC~5024, are well positioned on the horizontal branch for the NGC~5053's CMD, but shifted ``upward'' in case of NGC~5024's CMD (due to difference in the distance of these two globular clusters). In contrast, the one remaining extra-tidal RR Lyrae of NGC~5024 fits well to the horizontal branch of NGC~5024 (cyan point in Figure \ref{fig_groupA_cmd}).

To further evaluate the associations of these known RR Lyrae to the
two globular clusters, we compared their proper motions with the
measured proper motions of NGC~5024 and NGC~5053 in Figure
\ref{fig_groupA_pm}. The circles in this Figure denote the
``boundaries'' of the proper motions such that RR Lyrae located within
the circles are considered to have proper motions consistent with the
globular clusters. It is worth pointing out that proper motions of the
two globular clusters, as measured in \citet{gaia2018gc}, are close to
each other: the proper motions in Right Ascension are
$\mathrm{pmRA}_c^{5024}=-0.1466\pm0.0045\ \mathrm{(mas/year)}$ and
$\mathrm{pmRA}_c^{5053}=-0.3591\pm0.0071\ \mathrm{(mas/year)}$; while
in the Declination they are
$\mathrm{pmDE}_c^{5024}=-1.3514\pm0.0032\ \mathrm{(mas/year)}$ and
$\mathrm{pmDE}_c^{5053}=-1.2586\pm0.0048\ \mathrm{(mas/year)}$. Therefore,
it is difficult to associate  these RR Lyrae with either of the globular clusters based on proper motions alone. Figure \ref{fig_groupA_pm} reveals that this is indeed the case: proper motions of the known RR Lyrae in NGC~5024 are consistent with the proper motions of both globular clusters (left panels of Figure \ref{fig_groupA_pm})\footnote{There are few NGC~5024 RR Lyrae located outside the circles in left panels of Figure \ref{fig_groupA_pm}, indicating a possibility that they may not belong to NGC~5024. However they are all located within $\sim2.3\arcmin$ from the center of NGC~5024. The investigation of their memberships with NGC~5024 is not main scope of this paper, hence we will not study them in details further.}, and a similar situation holds for the RR Lyrae in NGC~5053 (right panels of Figure \ref{fig_groupA_pm}). For the five extra-tidal RR Lyrae identified in \citet{kundu2019}, their proper motions are fully consistent with either NGC~5024 or NGC~5053 (middle panels of Figure \ref{fig_groupA_pm}). This explains why \citet{kundu2019} would include the four RR Lyrae from NGC~5053 as the extra-tidal RR Lyrae for NGC~5024 based on the proper motion analysis.

Finally, we comment on the only extra-tidal RR Lyrae of NGC~5024, V48, which is located $12.81\arcmin$ away from NGC~5024. In \citet{kundu2019}, the adopted tidal radius of NGC~5024 is $18.37\arcmin$, hence $2/3$ of it is $12.25\arcmin$ which puts V48 at a borderline to be considered an extra-tidal RR Lyrae. In contrast, $2/3$ of the tidal radius adopted in this work is $15.20\arcmin$, then V48 will no longer be an extra-tidal RR Lyrae of NGC~5024. The cutoff of 2/3 of the tidal radius was based on the criterion defined in \citet{kundu2019}, at which the authors did not elaborate the reason for adopting such cutoff radius (also, see the discussion in Section 4). Other tidal radii, in arc-minutes, of NGC~5024 that can be found in the literature range from $14.79\pm7.19$ \citep{jordi2010}, 16.25 \citep{mvdm2005}, 16.91 \citep{kharchenko2013}, 21.85 \citep{peterson1975}, $21.87\pm0.53$ \citep{lehmann1997} and 22.48 \citep{trager1995}\footnote{Values taken from Table 4 of \citet{lehmann1997}.}. Therefore, depending on the adopted tidal radius, V48 could be either an extra-tidal RR Lyrae or not.

\subsection{Group B: Other Known RR Lyrae in the Vicinity}

\begin{figure*}
  %\epsscale{1.2}
  \plotone{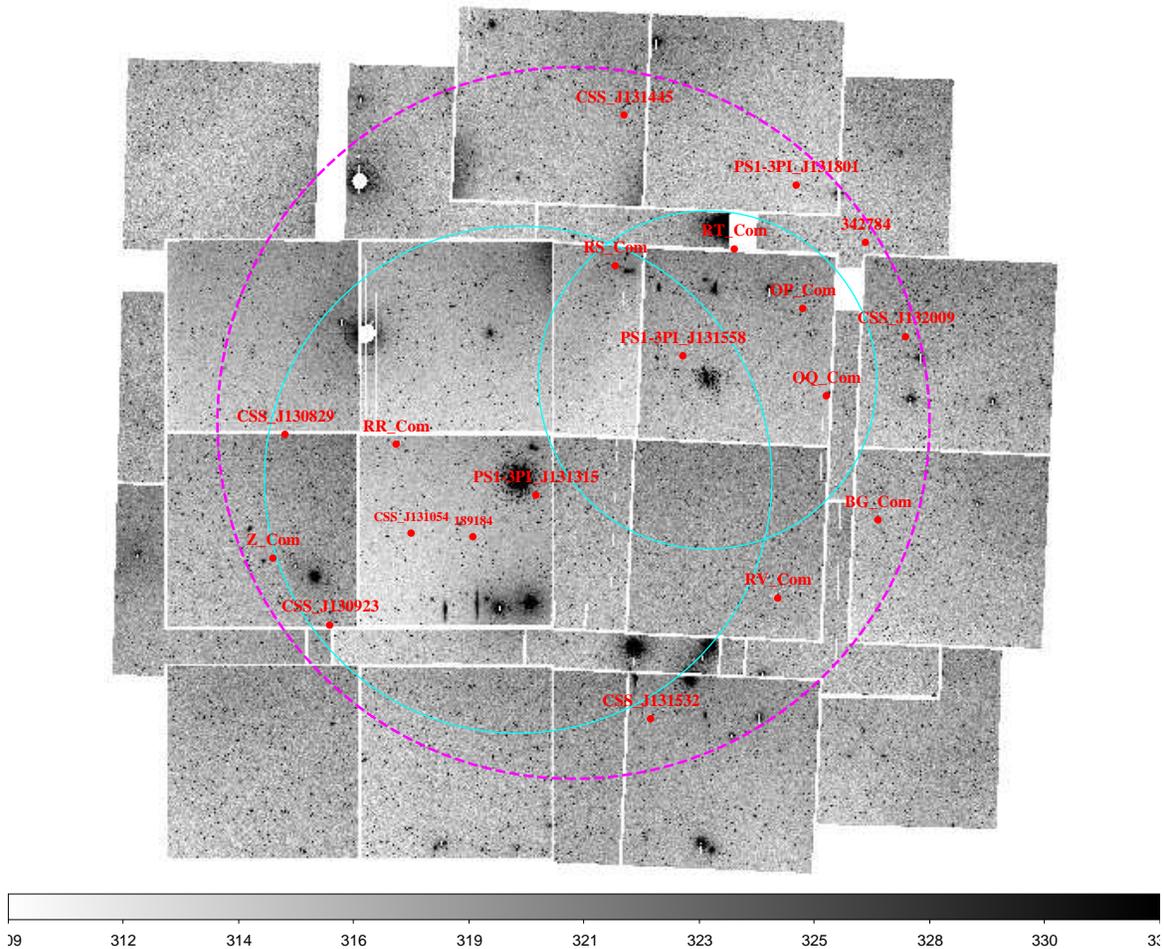}
  \caption{Locations of the 17 known RR Lyrae in ``Group B'' on the
    (inverted) ZTF mosaic  $r$-band reference images. Two cyan circles
    indicate the $3r_t$ radii for both globular clusters, while the dashed magenta circle represents the adopted search radius of $1.6$\,degree. For clarity, names for some of the RR Lyrae were shortened (see Table \ref{tab2}), and for the two RR Lyrae listed in the GaiaDR2RRL catalog only the last six digits were shown.}\label{fig_grouB_img}
\end{figure*}

\begin{figure*}
  \epsscale{1.2}
  \plotone{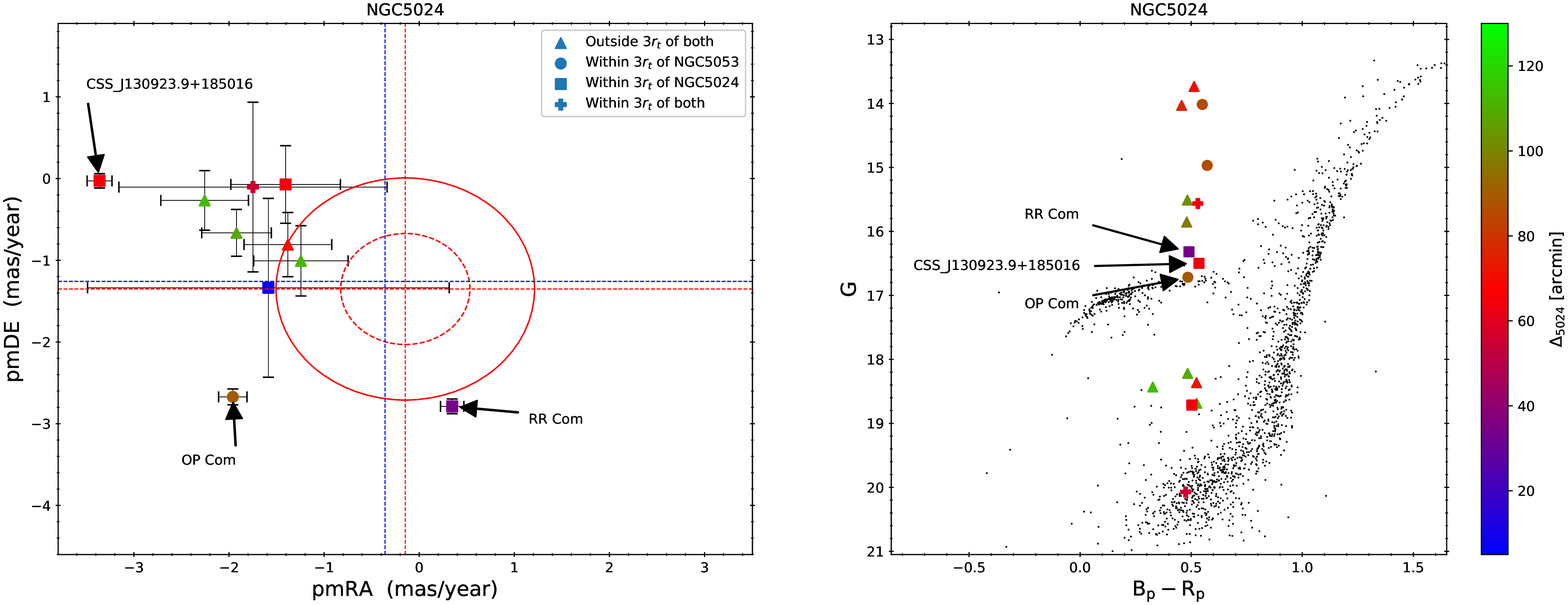}
  \plotone{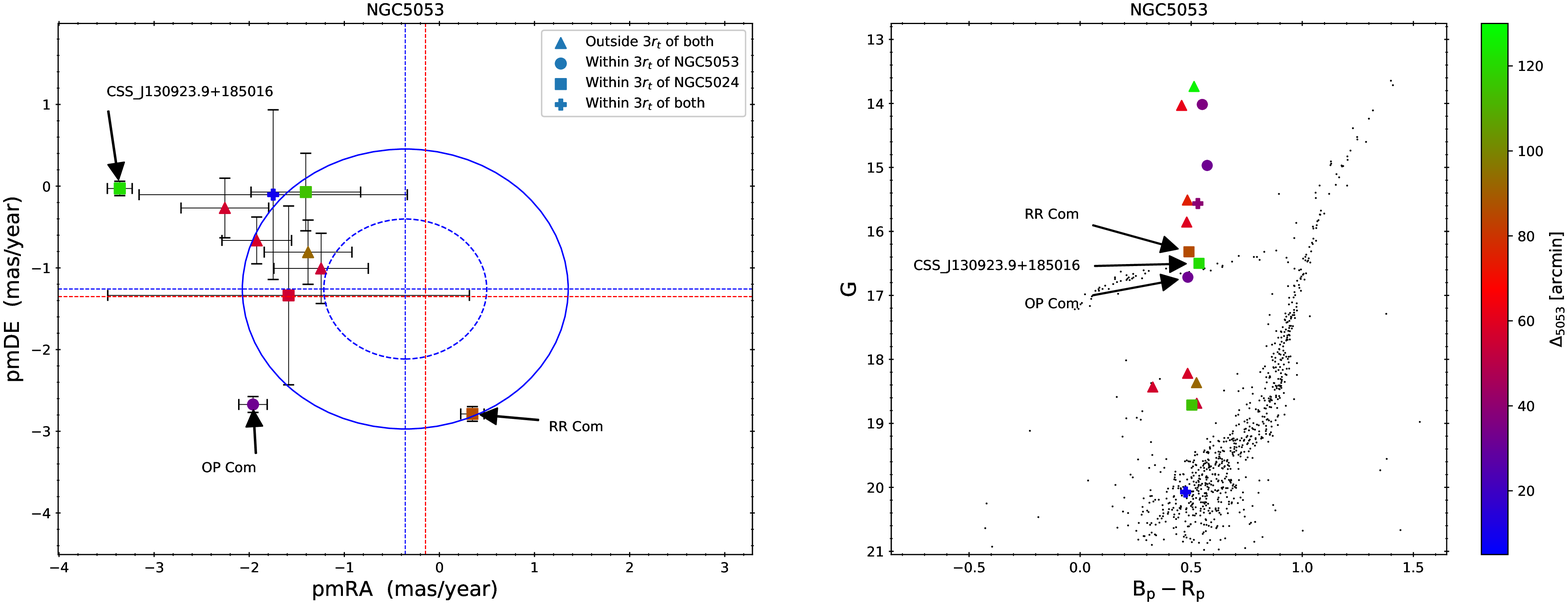}
  \caption{{\it Left panels:} Comparison of the proper motions for the known RR Lyrae in Group B, as listed in Table \ref{tab2}. These RR Lyrae were further divided into four sub-groups. Definitions of the circles and dashed lines are same as in Figure \ref{fig_groupA_pm}. {\it Right panels:} Positions of these RR Lyrae on CMD. The color bars represent the distance ($\Delta$, in arc-minute) to the center of either globular cluster.}\label{fig_grouB}
\end{figure*}

\begin{deluxetable*}{lccccccccc}
  %\tabletypesize{\scriptsize}
  \tabletypesize{\tiny}
  \tablecaption{Known RR Lyrae in the vicinity of NGC~5024 and NGC~5053 (Group B).\label{tab2}}
  \tablewidth{0pt}
  \tablehead{
    \colhead{VSX Name} &
    \colhead{$\alpha_{J2000}$} &
    \colhead{$\delta_{J2000}$} &
    \colhead{$\Delta_{5024}$} &
    \colhead{$\Delta_{5053}$} &
    \colhead{Gaia DR2 ID} &
    \colhead{$G$} &
  \colhead{$B_p-R_p$} &
  \colhead{pmRA} &
  \colhead{pmDE} }
  \startdata
  \cutinhead{Within the $3r_t$ of both clusters (filled plus signs in Figure \ref{fig_grouB})} \\
  PS1-3PI J131558\tablenotemark{a}& 198.99365 & 17.59436 & 55.55 & 9.32 & 3938679383472421760 & 20.073 & 0.475 & $-1.748\pm1.410$ & $-0.104\pm1.038$ \\
  RS Com & 198.66493 & 17.19691 & 63.35 & 39.62 & 3936953979146081792 & 15.564 & 0.530 & $-6.653\pm0.096$ & $-3.796\pm0.068$ \\
  \cutinhead{Within the $3r_t$ of NGC~5024 (filled squares in Figure \ref{fig_grouB})}
  PS1-3PI J131315\tablenotemark{a} & 198.31514 & 18.23614 & 6.33 & 55.74 & 3938773288637827712 & -99.99 & -99.99 & $-1.586\pm1.901$ & $-1.337\pm1.094$ \\
  $\cdots$                       & 198.02127 & 18.42893	& 19.66& 76.08 & 3938041533584189184 & -99.99\tablenotemark{b} & -99.99\tablenotemark{b} & -99.99 & -99.99 \\
  CSS J131054\tablenotemark{a} & 197.72836 & 18.41797 & 32.28 & 89.96 & 3938090393134378752 & -99.99 & -99.99 & $6.555\pm0.057$ & $-7.511\pm0.038$ \\
  RR Com & 197.64986 & 18.01958 & 34.28 & 85.72 & 3938004631227430144 & 16.318 & 0.490 & $0.347\pm0.123$ & $-2.788\pm0.090$ \\
  CSS J130829\tablenotemark{a} & 197.12431 & 17.98337 & 64.05 & 114.84 & 3937956660735121664 & 18.714 & 0.503 & $-1.405\pm0.576$ & $-0.072\pm0.475$ \\
  CSS J130923\tablenotemark{a} & 197.34989 & 18.83797 & 64.22 & 121.45 & 3938306378448284672 & 16.499 & 0.535 & $-3.361\pm0.131$ & $-0.028\pm0.088$ \\
  \cutinhead{Within the $3r_t$ of NGC~5053 (filled circles in Figure \ref{fig_grouB})}
  OP Com & 199.55205 & 17.36733 & 89.52 & 32.10 & 3938423098479656960 & 16.716 & 0.485 & $-1.960\pm0.149$ & $-2.672\pm0.097$ \\
  OQ Com & 199.67464 & 17.75669 & 86.06 & 32.28 & 3938510097337185280 & 14.969 & 0.572 & $1.007\pm0.079$ & $-5.827\pm0.067$ \\
  RT Com & 199.22275 & 17.10860 & 85.22 & 36.05 & 3936916973707946752 & 14.016 & 0.550 & $5.249\pm0.053$ & $-13.564\pm0.042$ \\
  \cutinhead{Not within the $3r_t$ of either clusters (filled triangles in Figure \ref{fig_grouB})}
  Z Com & 197.07614 & 18.54048 & 69.41 & 126.61 & 3938269613528220672 & 13.735 & 0.513 & $3.182\pm0.049$ & $-19.942\pm0.046$ \\
  RV Com & 199.47250 & 18.67250 & 76.92 & 61.83 & 3938893311498825984 & 14.031 & 0.457 & $-11.983\pm0.058$ & $-5.711\pm0.050$ \\
  BG Com & 199.93618 & 18.30730 & 97.57 & 59.44 & 3938639839708973056 & 15.854 & 0.480 & $-5.195\pm0.100$ & $-1.275\pm0.080$ \\
  CSS J131532\tablenotemark{a} & 198.88418 & 19.22940 & 73.73 & 92.67 & 3939689422341157120 & 18.364 & 0.524 & $-1.381\pm0.461$ & $-0.808\pm0.393$ \\
  CSS J131445\tablenotemark{a} & 198.68950 & 16.51934 & 102.37 & 74.90 & 3936774724391121280 & 15.510 & 0.483 & $-4.652\pm0.089$ & $-3.667\pm0.062$ \\
  CSS J132009\tablenotemark{a} & 200.03928 & 17.48136 & 111.25 & 54.59 & 3938440072190455296 & 18.686 & 0.525 & $-1.244\pm0.496$ & $-1.007\pm0.430$ \\
  $\cdots$                         & 199.83771 & 17.06256	& 113.36 & 56.45 & 3746246947188342784 & 18.432 & 0.327	& $-2.256\pm0.461$ & $-0.267\pm0.365$ \\
  PS1-3PI J131801\tablenotemark{a} & 199.50486 & 16.81444 & 109.17 & 57.70 & 3936893742228790528 & 18.218 & 0.484 & $-1.921\pm0.366$ & $-0.664\pm0.286$ 
  \enddata
  \tablenotetext{a}{PS1-3PI J131558 = PS1-3PI J131558.47+173539.6; PS1-3PI J131315 = PS1-3PI J131315.63+181410.1; PS1-3PI J131801 = PS1-3PI J131801.16+164851.9; CSS J130829 = CSS J130829.7+175900; CSS J130923 = CSS J130923.9+185016; CSS J131054 = CSS J131054.8+182504; CSS J131532 = CSS J131532.1+191346; CSS J131445 = CSS J131445.4+163109; CSS J132009 = CSS J132009.3+172853.}
  \tablenotetext{b}{The Gaia DR2 main catalog listed $G=19.480\pm0.020$ and $B_P-R_p=1.392$ for this RR Lyrae.}
  \tablecomments{The meanings of each columns are same as in Table \ref{tab1}.}
\end{deluxetable*}

For the remaining 22 RR Lyrae, 17 and 5 of them are known RR Lyrae
from the VSX and GaiaDR2RRL catalogs, respectively. Among the 5 RR
Lyrae from the GaiaDR2RRL catalog, 3 of them are located within
$2\arcmin$ of the center of NGC~5024\footnote{The Gaia DR2 ID for
  these 3 RR Lyrae are: 3938022017256004352, 3938022085975309952 and
  3938022463930328960.}. Two of them have $G>19.7$\,mag hence they
could be the background stars, and the RR Lyrae with Gaia DR2 ID of
3938022017256004352 has a $G$-band magnitude of
$16.538\pm0.001$\,mag. This RR Lyrae could be a new member of
NGC~5024, but no proper motion information available in the Gaia DR2
main catalog. Nevertheless, we excluded these three RR Lyrae in this
work. The locations of other 19 RR Lyrae with respect to NGC~5024 and
NGC~5053 are shown in Figure \ref{fig_grouB_img}. These RR Lyrae were
further grouped into four sub-groups based on their locations relative
to the two globular clusters, as summarized in Table \ref{tab2}. The
comparisons of their proper motions to those of NGC~5024 and NGC~5053, as well as their positions on the CMD, are displayed in Figure \ref{fig_grouB}. Two RR Lyrae, PS1-3PI J131315.63+181410.1 (or PS1-3PI J131315 in Table \ref{tab2}) and Gaia DR2 ID 3938041533584189184, were found to be mis-identified RR Lyrae (see Appendix B for more details).
 
Seven RR Lyrae appeared to have proper motions consistent with either
of the globular clusters as shown in the left panels of Figure
\ref{fig_grouB}, and they also satisfied the proper motion criterion
given in \citet{kundu2019}. However, these seven RR Lyrae are all
fainter than $G\sim 18$\,mag in the CMD\footnote{In case of PS1-3PI
  J131315.63+181410.1, there is no $G$-band intensity mean magnitude
  given in the {\tt gaiadr2.vary\_rrlyrae} Table. Nevertheless the
  Gaia DR2 main catalog listed a value of $G=20.222\pm0.007$\,mag for
  this RR Lyrae.}, as presented in right panel of Figure
\ref{fig_grouB}, hence they  most likely belong to the background
population of RR Lyrae in the Galactic halo. In contrast, three RR
Lyrae are located in the magnitude range of $16.0 < G < 17.5$, enclosing the expected magnitudes of the horizontal branch stars in NGC~5024 and NGC~5053. CSS J130923.9+185016 is located right on top of the horizontal branch of NGC~5053 but it is closer to NGC~5024 on the sky. Similarly, OP Com seems to match with the horizontal branch of
NGC~5024 but it is located near NGC~5053. Proper motions of these two RR Lyrae exhibit a large deviation relative to either of the globular clusters and hence they do not appear to be members or  extra-tidal RR Lyrae of these two clusters. The most promising candidate of the extra-tidal RR Lyrae is RR Com, which is located within  $3r_t$ of NGC~5024. The mean brightness of this RR Lyrae is $G=16.3176\pm0.0002$\,mag, about $\sim0.4$\,mag brighter than the horizontal branch of NGC~5024 (see upper-right panel of Figure \ref{fig_grouB}), and similar to the RR Lyrae V60 in NGC~5024 (with $G=16.3854\pm0.0009$\,mag). In fact, RR Com was picked up in \citet{kundu2019} as a potential extra-tidal RR Lyrae for NGC~5024 but rejected based on the proper motion criterion.

\section{Searching for new RR Lyrae}

Since none of the known RR Lyrae within our search area of $\sim8$\,deg$^2$ were found to be definitive extra-tidal RR Lyrae of NGC~5024 and NGC~5053, we attempt to search for new RR Lyrae within the same area in this Section. Given that RR Lyrae are high amplitude variable stars, and the expected extra-tidal RR Lyrae would have similar brightness as those in the NGC~5024 and NGC~5053 with $G\sim16.7$\,mag (the averaged value for RR Lyrae listed in Table \ref{tab1}), any such new extra-tidal RR Lyrae should already have been detected from  time-domain all-sky surveys such as Pan-STARRS1 \citep[as done in][]{sesar2017} or Gaia \citep{clementini2019} that can reach to a depth fainter than $\sim20$\,mag. The fact that we did not find any extra-tidal RR Lyrae, based on the known RR Lyrae, in the previous Sections indicates that most likely we will not find any new RR Lyrae around $G\sim16.7$\,mag located within our search area.   

\subsection{Selecting Candidates from Gaia DR2}

We began the search for new RR Lyrae using the Gaia DR2 main catalog. We selected $664$ stars that located within the search area of $\sim8$\,deg$^2$ defined in the previous Section, as well as satisfied the conditions  $16.0 < G < 17.5$ and $0.2 < (B_P - R_P) < 0.8$ (roughly bracketed the positions of expected RR Lyrae in the CMD). We further excluded those stars located within $2/3$ of the tidal radius of either globular clusters, leaving $291$ stars in our sample. Finally, we applied the proper motion criterion from \citet{kundu2019} in either the right ascension or declination directions, but not both,  leaving $79$ stars (including RR Com) to be examined with  ZTF light curves in the next subsection. If we enforce the \citet{kundu2019} proper motion criterion to be satisfied in both directions, this would leave only $9$ stars in the sample.   

\subsection{Light Curves from ZTF}

The Zwicky Transient Facility (ZTF, operating 2018--2020) is a dedicated time domain wide-field synoptic sky survey aimed to explore the transient universe. ZTF utilizes the Palomar 48-inch Samuel Oschin Schmidt telescope, together with a new mosaic CCD camera, that provides a field-of-view of 47 squared degree to observe the Northern sky in customized $gri$ filters. Further details regarding ZTF can be found in \citet{bel19,gra19,dec20} and will not be repeated here. Imaging data taken from ZTF were processed with a dedicated reduction pipeline, as detailed in \citet{mas19}; the final data products included reduced images and catalogs based on PSF (point-spread function) photometry.

\begin{figure*}
  \epsscale{1.1}
  \plottwo{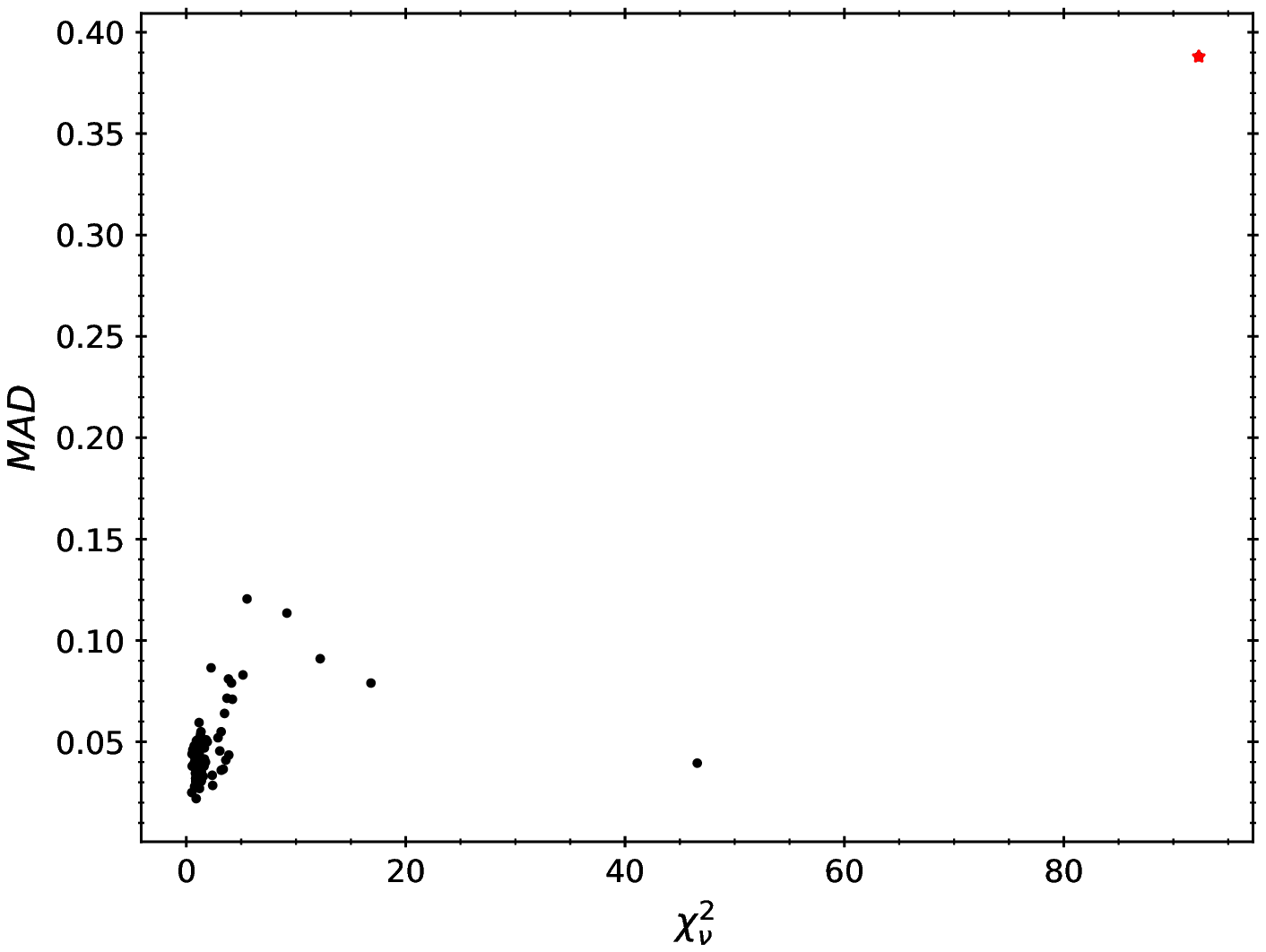}{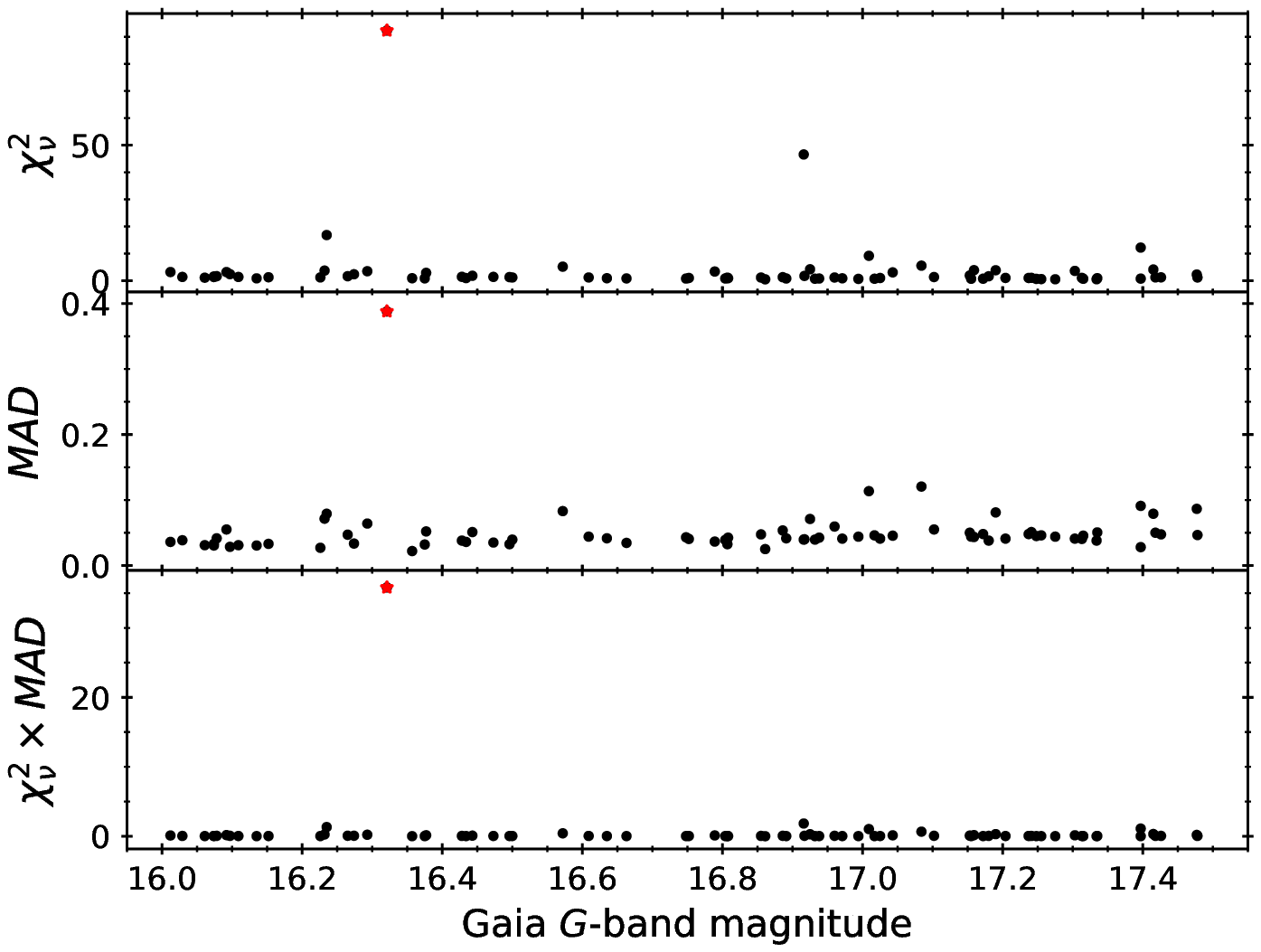}
  \caption{{\it Left panel:} $MAD$ vs. $\chi^2_\nu$ values for the 79 Gaia stars selected in Section 3.1 based on the $gri$-band ZTF light curves. {\it Right panel}: The $\chi^2_\nu$, $MAD$ and $\chi^2_\nu \times MAD$ values as a function of Gaia $G$-band magnitudes. The red starry symbols in all panels represent RR Com.}\label{fig_chimad}
\end{figure*}

Light curves of the $79$ stars identified in previous subsection were
extracted from the ZTF's PSF catalogs spanning from November
2017\footnote{Part of the data were taken during the ZTF commissioning
  phase from November 2017 to March 2018.} to January 2020. The number
of data points per light curves, in the format of
maximum/median/minimum, are: 13/110/202 in the $g$-band; 35/172/318 in
the $r$-band; and 5/39/78 in the $i$-band. Following \citet{yang2010},
we calculated the $\chi^2_\nu$ values for each star relative to a
constant model using all $gri$-band data:

\begin{figure*}
  \epsscale{1.1}
  \plottwo{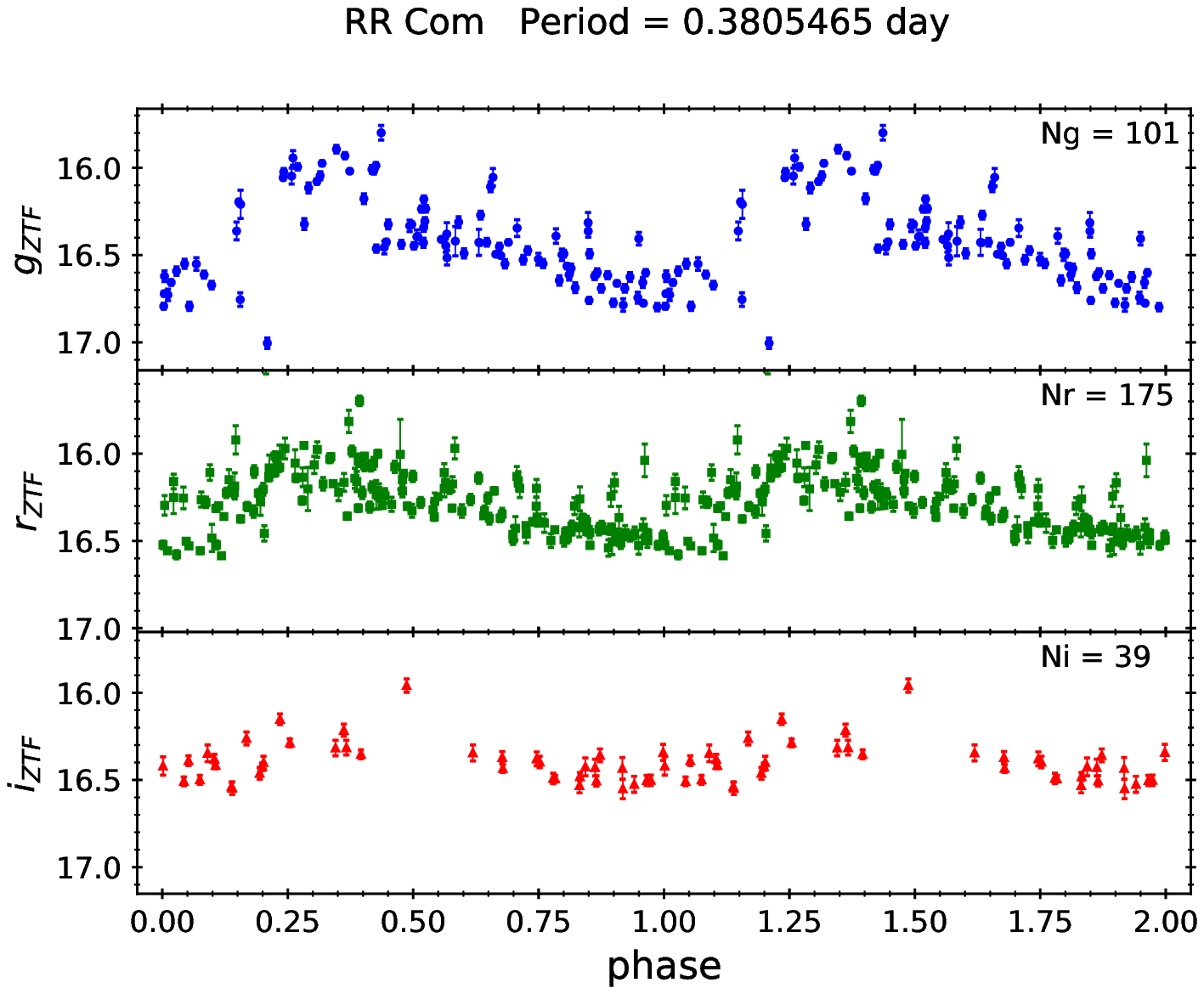}{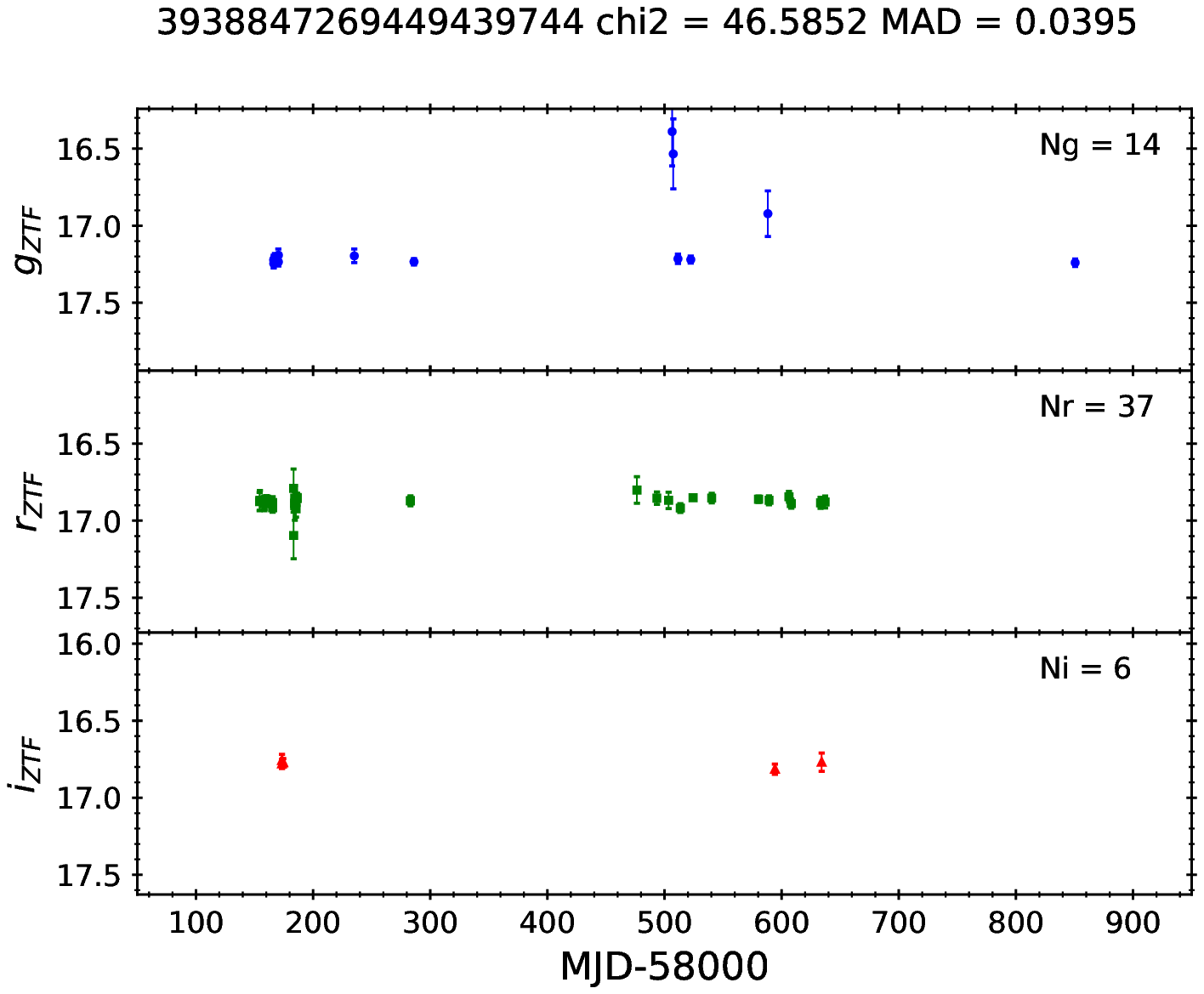}
  \caption{{\it Left panel:} ZTF light curves for RR Com, which is a known double mode pulsator \citep{poleski2014} that simultaneously pulsating in two periods. The ZTF light curves were folded with single period adopted from the VSX Catalog, hence the phased light curves appeared to exhibit scatters. {\it Right panel:} ZTF light curves for the star with the second highest $s \equiv \chi^2_\nu \times MAD$ value, which were affected by few outliers in the light curves. Nevertheless, it is clear that this star does not exhibit RR Lyrae-like light curves. If the obvious outliers were removed, the $\chi^2_\nu$ and $MAD$ values were reduced to $4.1286$ and $0.0340$, respectively, and this star is no longer belong to the second highest $s$ value in the sample. $N$ is the number of data points in each light curves.}\label{fig_top2ztf}
  \end{figure*}

\begin{eqnarray}
  \chi^2_\nu & = & \frac{1}{N} \left[ \sum_{j=1}^{N_g} \frac{(g_j-\bar{g})^2}{\sigma_{g,j}^2} +  \sum_{j=1}^{N_r} \frac{(r_j-\bar{r})^2}{\sigma_{r,j}^2} +  \sum_{j=1}^{N_j} \frac{(i_j-\bar{i})^2}{\sigma_{i,j}^2} \right], \nonumber
\end{eqnarray}

\noindent where $N=N_g + N_r + N_i-3$. To guard against outliers, we also calculated the $MAD$ (median absolute deviation) values for the same sets of light curves:

\begin{eqnarray}
  MAD & = & \sum_{x=\{g,r,i\}} \mathrm{median}\left( |x_j-\tilde{x}|\right), \nonumber
\end{eqnarray}

\noindent where $\tilde{x}$ is the median value of array $x$. The left
panel of Figure \ref{fig_chimad} presents the values of $MAD$
vs.\ $\chi^2_\nu$, where some large values of $\chi^2_\nu$ seem to be
affected by outliers. The product of these two values, $\chi^2_\nu
\times MAD \equiv s$, appears to be a good metric to distinguish
between  variable and non-variable stars, as shown in bottom right
panel of Figure \ref{fig_chimad}. The only known RR Lyrae among the 79
stars, RR Com,  clearly stands out in this figure. Light curves for
the two stars (including RR Com) with the highest values of $s$ are
presented in Figure \ref{fig_top2ztf}. Undoubtedly, the star in the
right panel of Figure \ref{fig_top2ztf} is not a RR Lyrae (nor a
variable star of any sort).

\begin{figure*}
  \epsscale{1.1}
  \plottwo{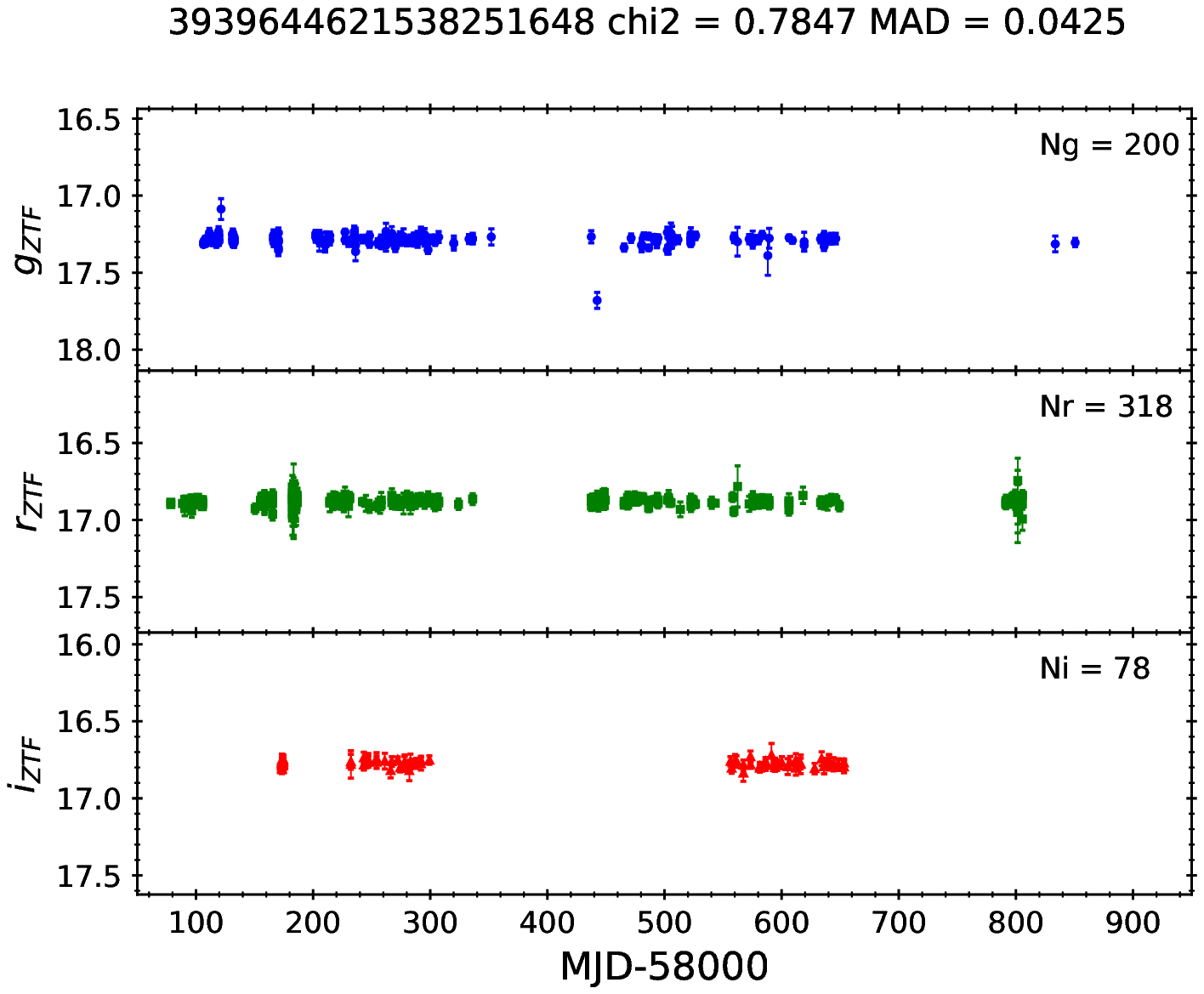}{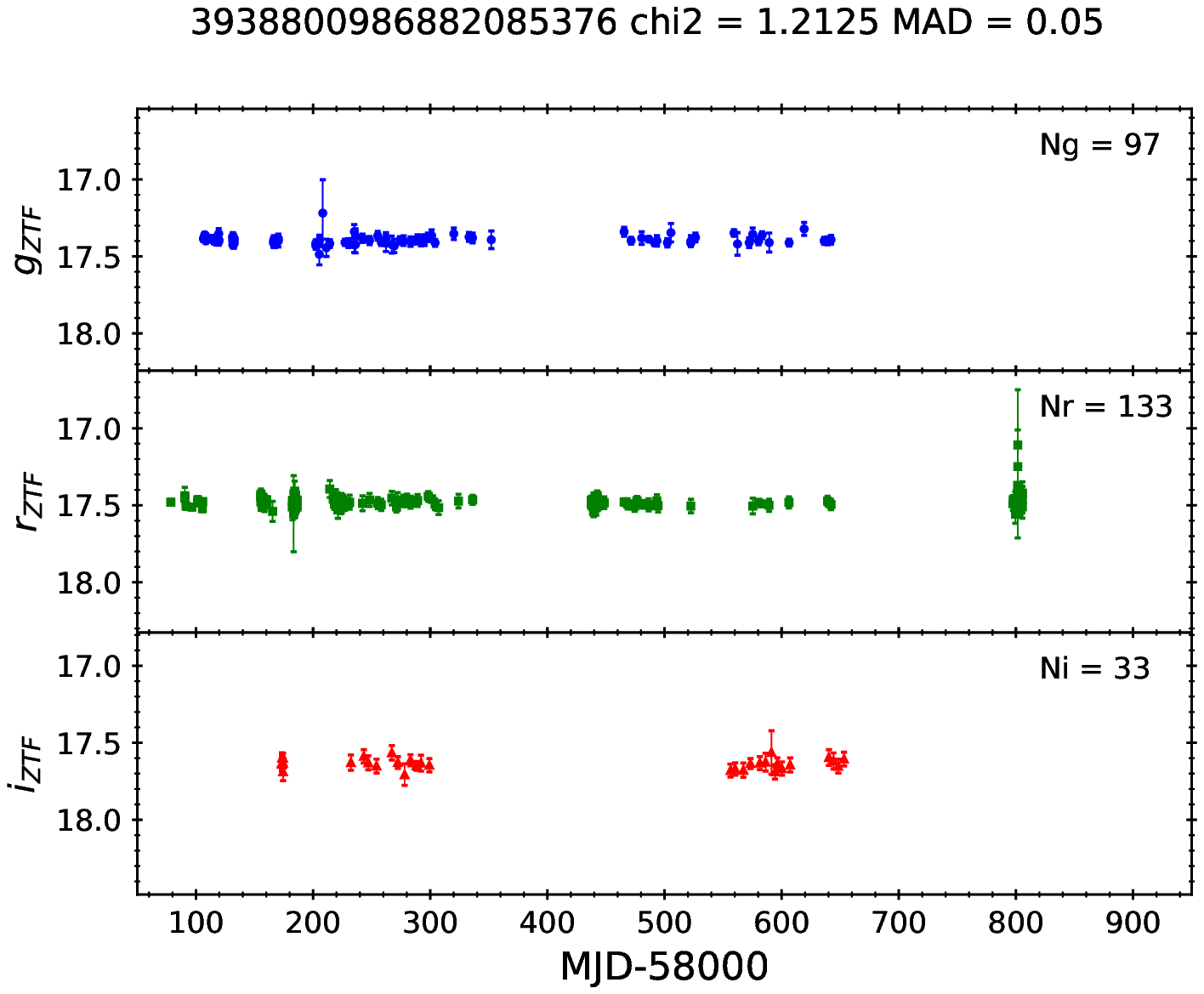}
  \plottwo{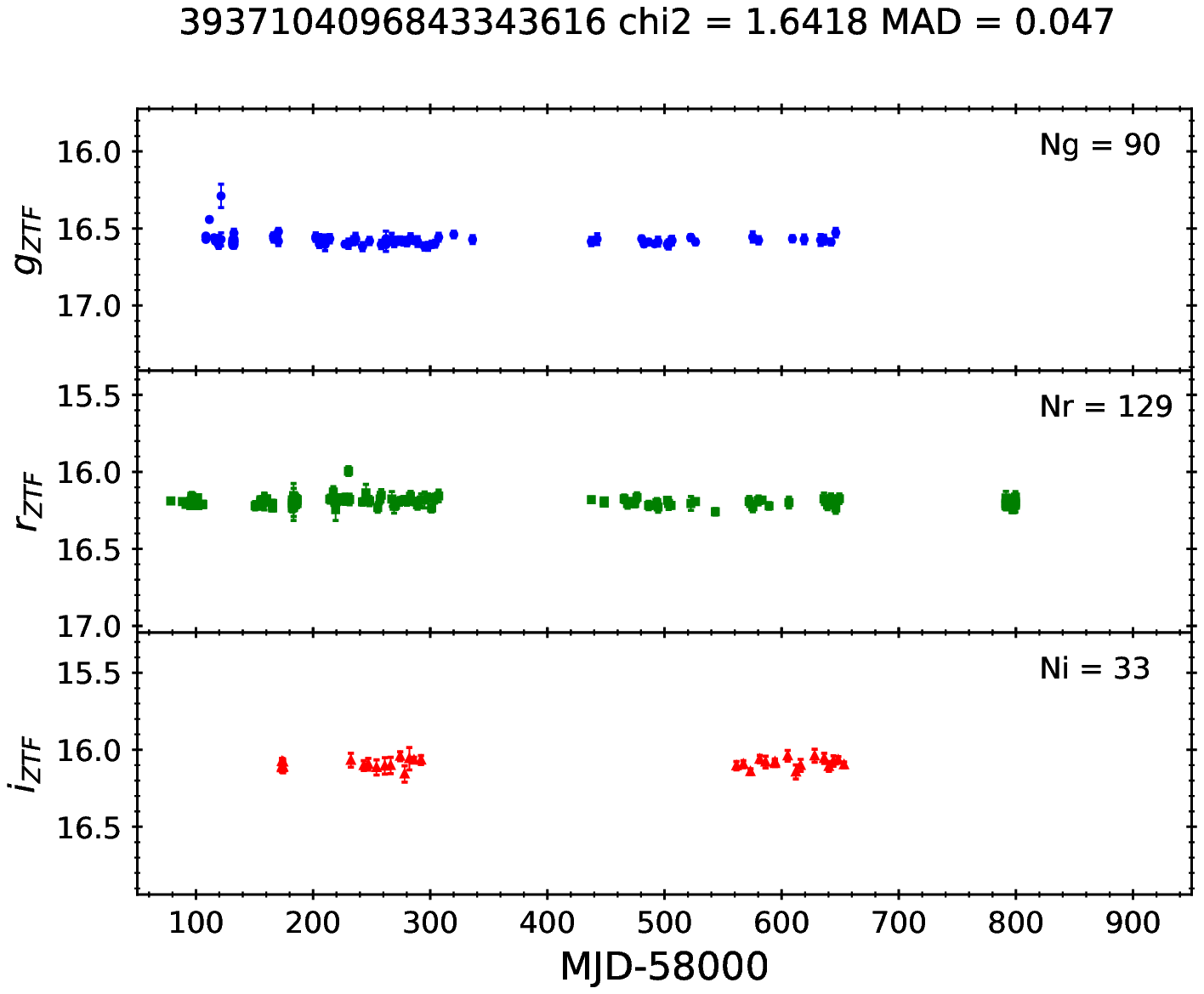}{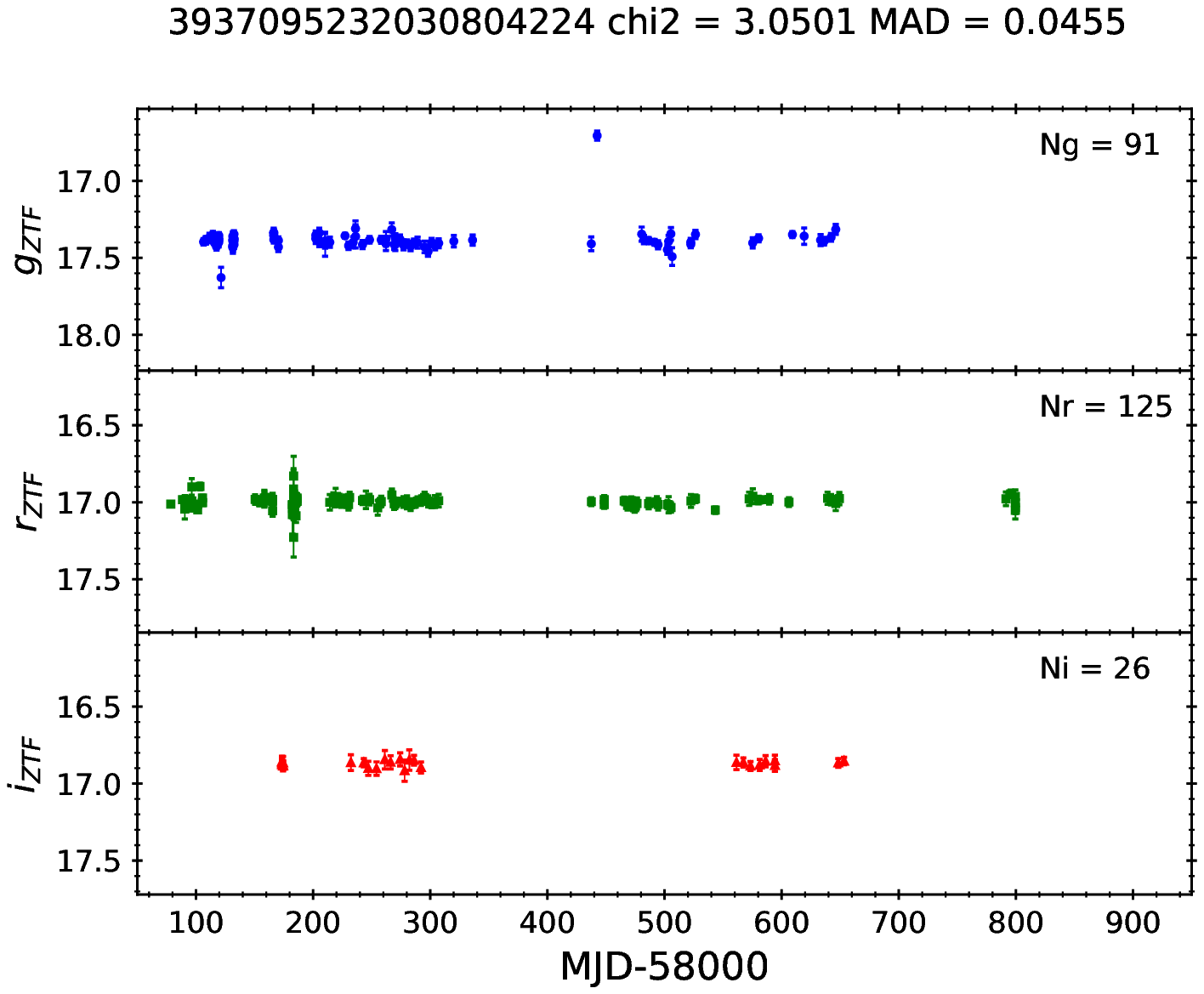}
  \caption{ZTF light curves for randomly selected stars with $s<0.17$, where $s\equiv \chi^2_\nu \times MAD$. $N$ is the number of data points in each light curves.}\label{fig_nonvar}
\end{figure*}

\begin{figure*}
  \epsscale{1.1}
  \plottwo{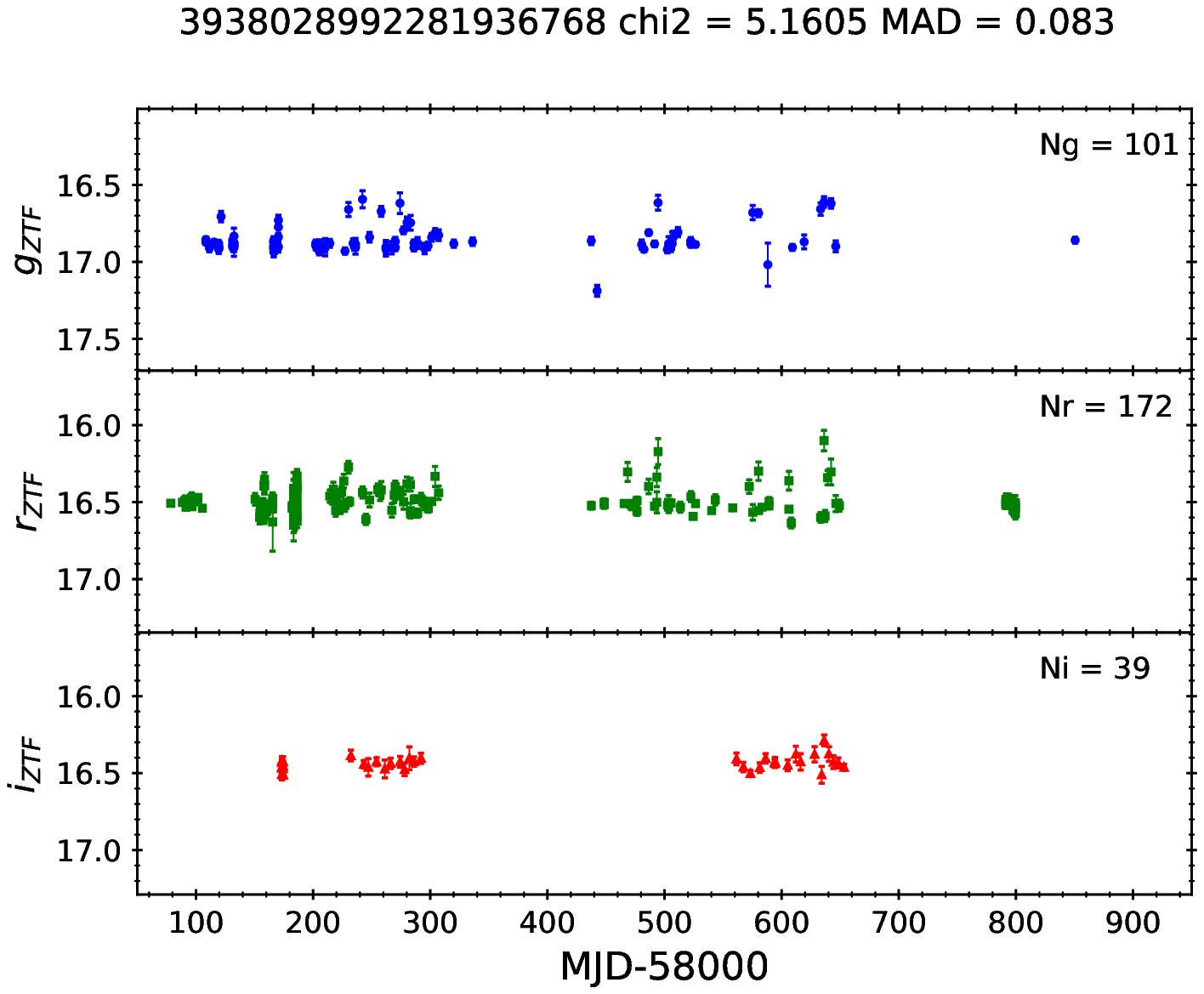}{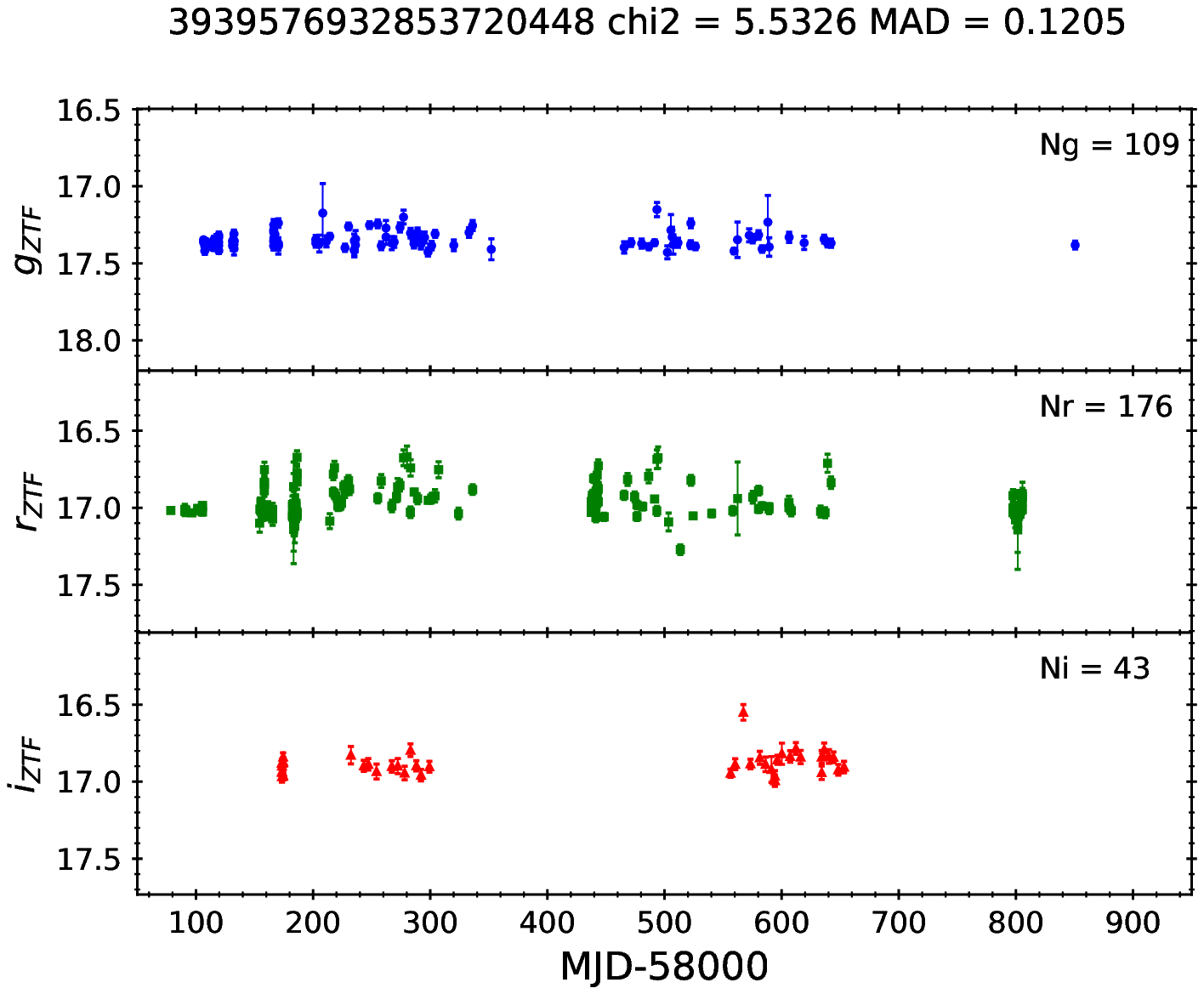}
  \plottwo{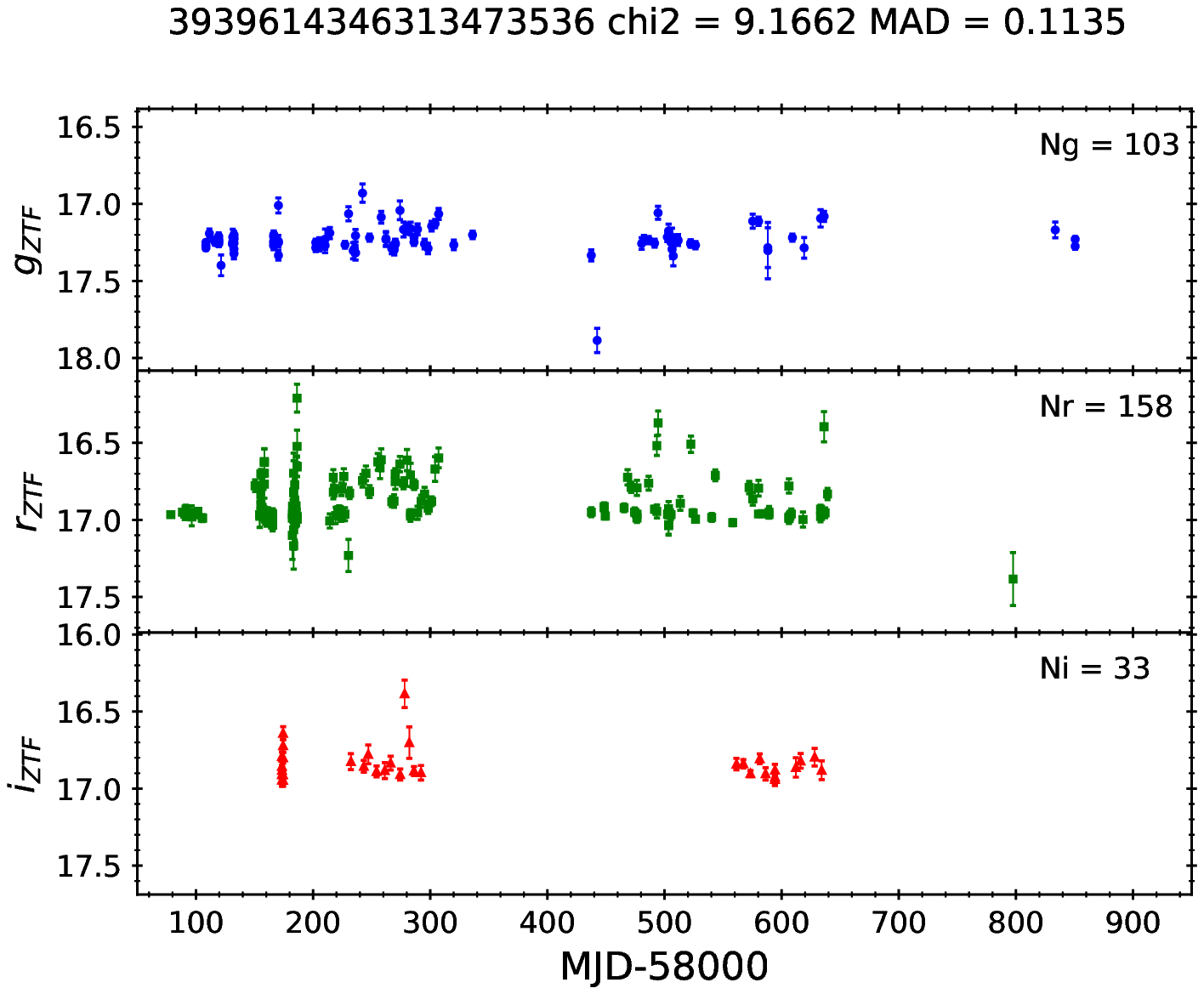}{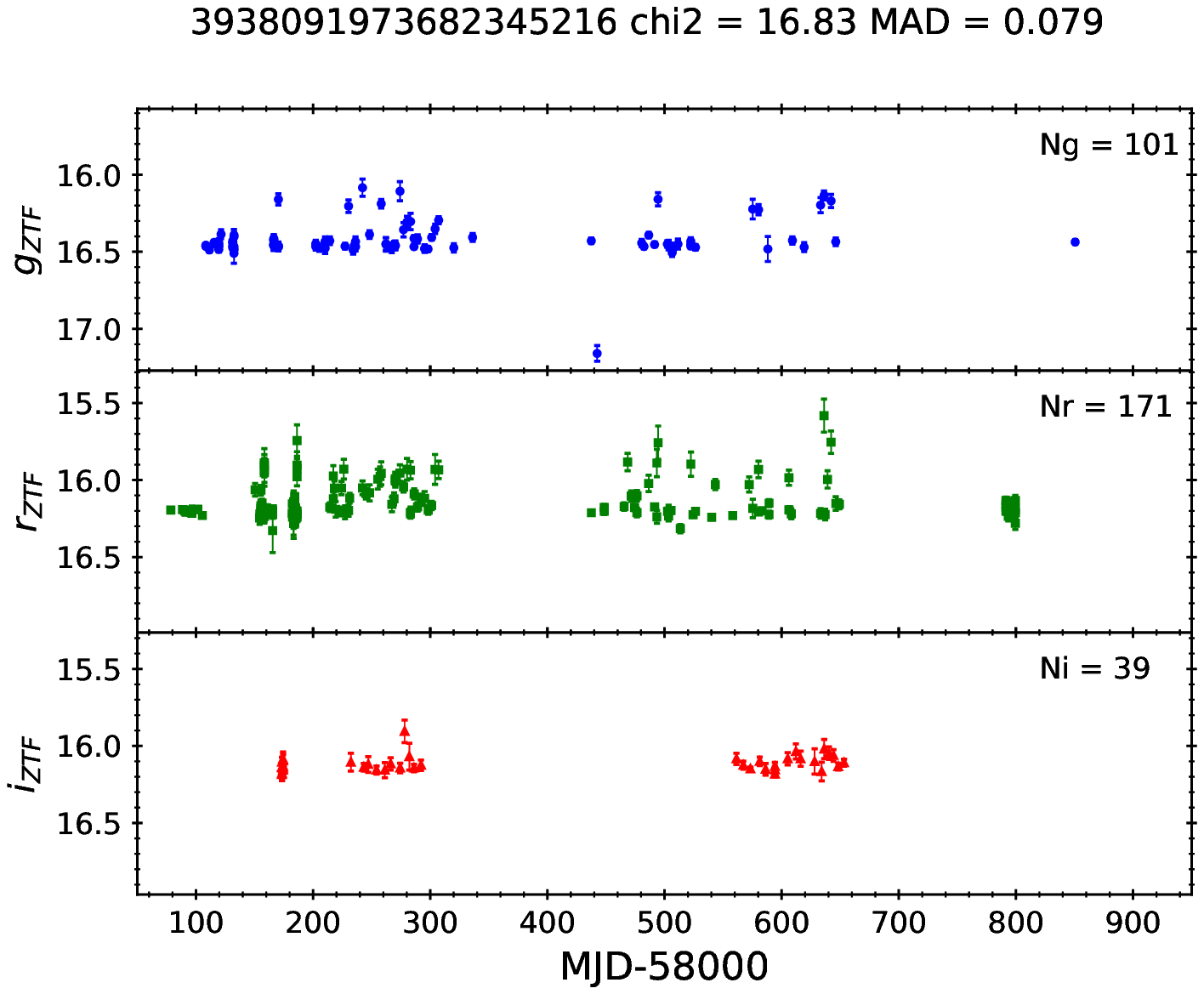}
  \caption{ZTF light curves for stars with $s>0.17$, where $s\equiv \chi^2_\nu \times MAD$, that exhibit variability but no credible period can be detected. $N$ is the number of data points in each light curves.}\label{fig_varcand}
\end{figure*}

Visually inspecting these light curves reveals that  those light
curves with $s<0.17$ do not shown signs of variability. There were 65
stars with $s<0.17$ in the sample, including the 9 stars that
satisfied the proper motion criterion in \citet{kundu2019} mentioned
in the previous subsection.  A few example light curves are presented
in Figure \ref{fig_nonvar}. We searched for a periodicity  on the rest of stars with $s>0.17$, excluding RR Com, using the Lomb-Scargle periodogram implemented in the Astropy package \citep{astropy2013,astropy2018}. The periodicity search  was done on the $r$-band light curves (as this band has the most data points) within the period range of 0.2 to 1.2\,days, appropriate for the possible periods of RR Lyrae. Among about a dozen stars with $s>0.17$, the majority of them do not show any periodicity, and there were four stars displaying some variability but without a convincing periodicity. Their ZTF light curves are displayed in Figure \ref{fig_varcand}, and none of them is an RR Lyrae.

\section{Conclusion}

In this work, we re-evaluated the claim of \citet{kundu2019} that there are five extra-tidal RR Lyrae associated with globular cluster NGC~5024. Four of these RR Lyrae were known members of a nearby globular cluster NGC~5053. The remaining extra-tidal RR Lyrae, V48 of NGC~5024, could either be an extra-tidal RR Lyrae of NGC~5024 or not -- depending on the adopted tidal radius of NGC~5024. One of the criteria employed in \citet{kundu2019} is such that extra-tidal RR Lyrae should be located outside the $2/3$ of the tidal radius for a given globular cluster. Other similar work in literature, however, adopted the criterion of one tidal radius to select the extra-tidal RR Lyrae \citep{kunder2018,minniti2018}. If the criterion of one tidal radius is adopted, then V48 will no longer be an extra-tidal RR Lyrae of NGC~5024.

Besides the RR Lyrae that were known members of NGC~5024 and NGC~5053 from Clement's catalog, we compiled a list of known RR Lyrae within an area of $\sim8$\,deg$^2$ from literature. Using the similar selection criteria as in \citet{kundu2019}, together with the Gaia DR2 data, none of these known RR Lyrae were found to be extra-tidal RR Lyrae of either globular clusters. Finally, using Gaia DR2 data, we selected stars within our search area but outside the 2/3 of the tidal radii of both globular clusters that fall in the range of $16.0 < G <17.5$ and $0.2 < (B_P - R_P) < 0.8$, as well as satisfied the proper motion criterion from \citet{kundu2019} in either directions. A further variability and periodicity analysis of these stars with the ZTF light curves data revealed that none of them were new RR Lyrae. Therefore, we have concluded there were no extra-tidal RR Lyrae for either NGC~5024 or NGC~5053 within our search area of $\sim8$\,deg$^2$ that covering both globular clusters.

\acknowledgments

We thank the useful discussions and comments from T. de Boer, P. Mroz and an anonymous referee to improve the manuscript. We thank the funding from Ministry of Science and Technology (Taiwan) under the contract 107-2119-M-008-014-MY2, 107-2119-M-008 012 and 108-2628-M-007-005-RSP.

Based on observations obtained with the Samuel Oschin Telescope 48-inch and the 60-inch Telescope at the Palomar Observatory as part of the Zwicky Transient Facility project. Major funding has been provided by the U.S. National Science Foundation under Grant No. AST-1440341 and by the ZTF partner institutions: the California Institute of Technology, the Oskar Klein Centre, the Weizmann Institute of Science, the University of Maryland, the University of Washington, Deutsches Elektronen-Synchrotron, the University of Wisconsin-Milwaukee, and the TANGO Program of the University System of Taiwan. 

This work has made use of data from the European Space Agency (ESA) mission {\it Gaia} (\url{https://www.cosmos.esa.int/gaia}), processed by the {\it Gaia} Data Processing and Analysis Consortium (DPAC, \url{https://www.cosmos.esa.int/web/gaia/dpac/consortium}). Funding for the DPAC has been provided by national institutions, in particular the institutions participating in the {\it Gaia} Multilateral Agreement.

This research has made use of the SIMBAD database and the VizieR catalogue access tool, operated at CDS, Strasbourg, France. This research has made use of the International Variable Star Index (VSX) database, operated at AAVSO, Cambridge, Massachusetts, USA.

This research made use of Astropy,\footnote{http://www.astropy.org} a community-developed core Python package for Astronomy \citep{astropy2013, astropy2018}.

\facility{Gaia, PO:1.2m}

\software{Astropy package \citep{astropy2013,astropy2018}, Astroquery \citep{astroquery2019}}

\appendix

\section{Constructing the Proper Motion and Color-Magnitude Diagrams for the Clusters}

\begin{figure*}
  \plottwo{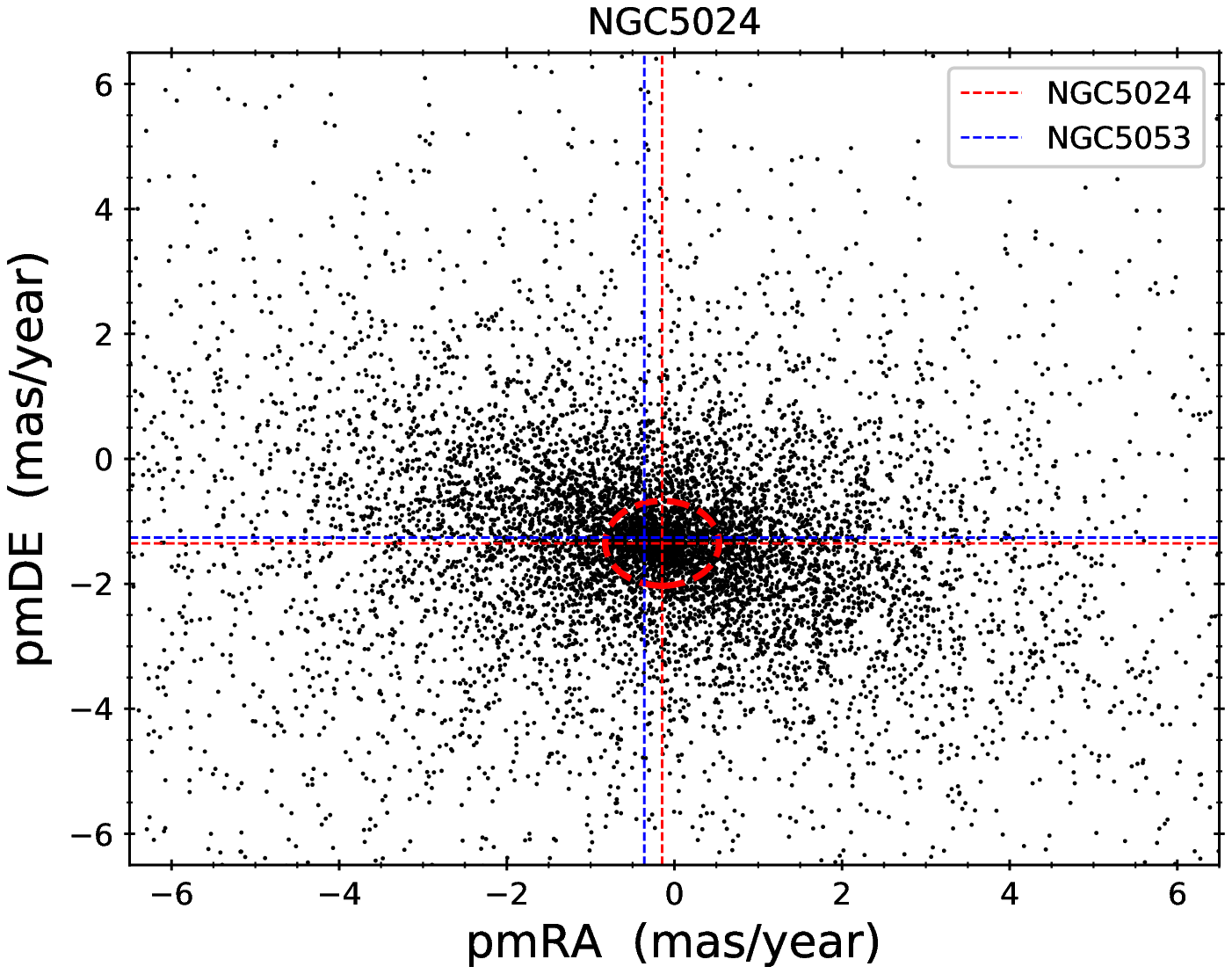}{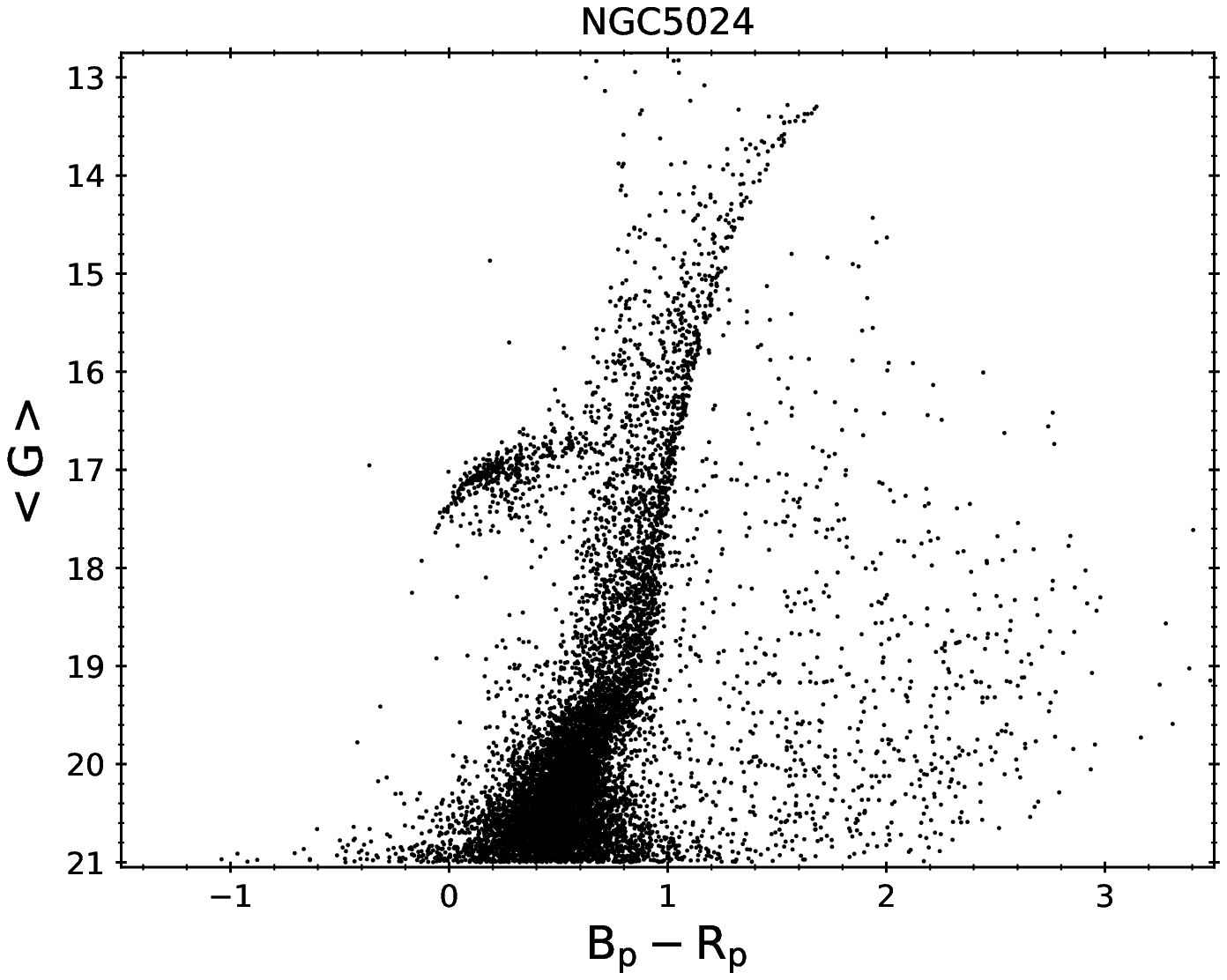}
  \plottwo{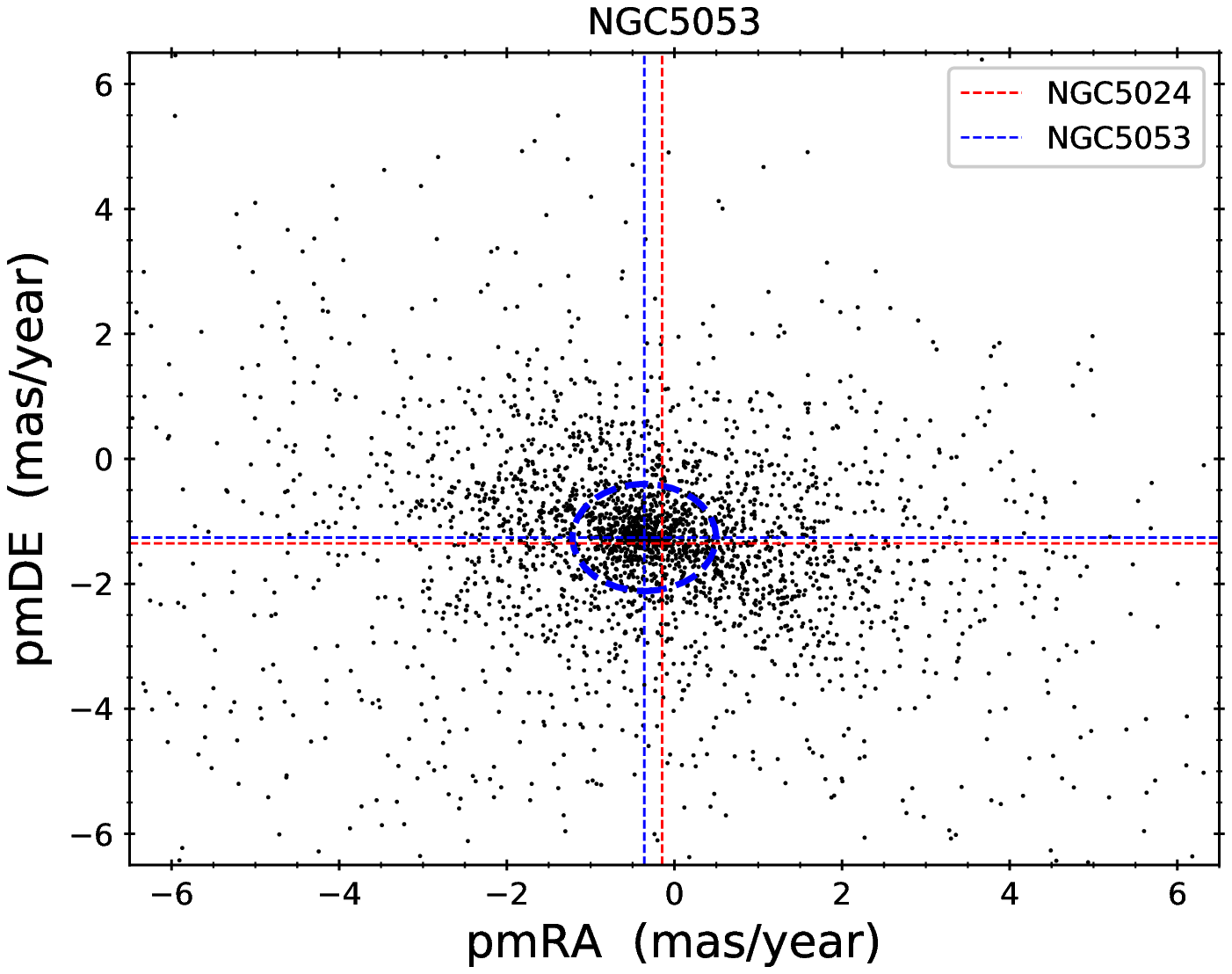}{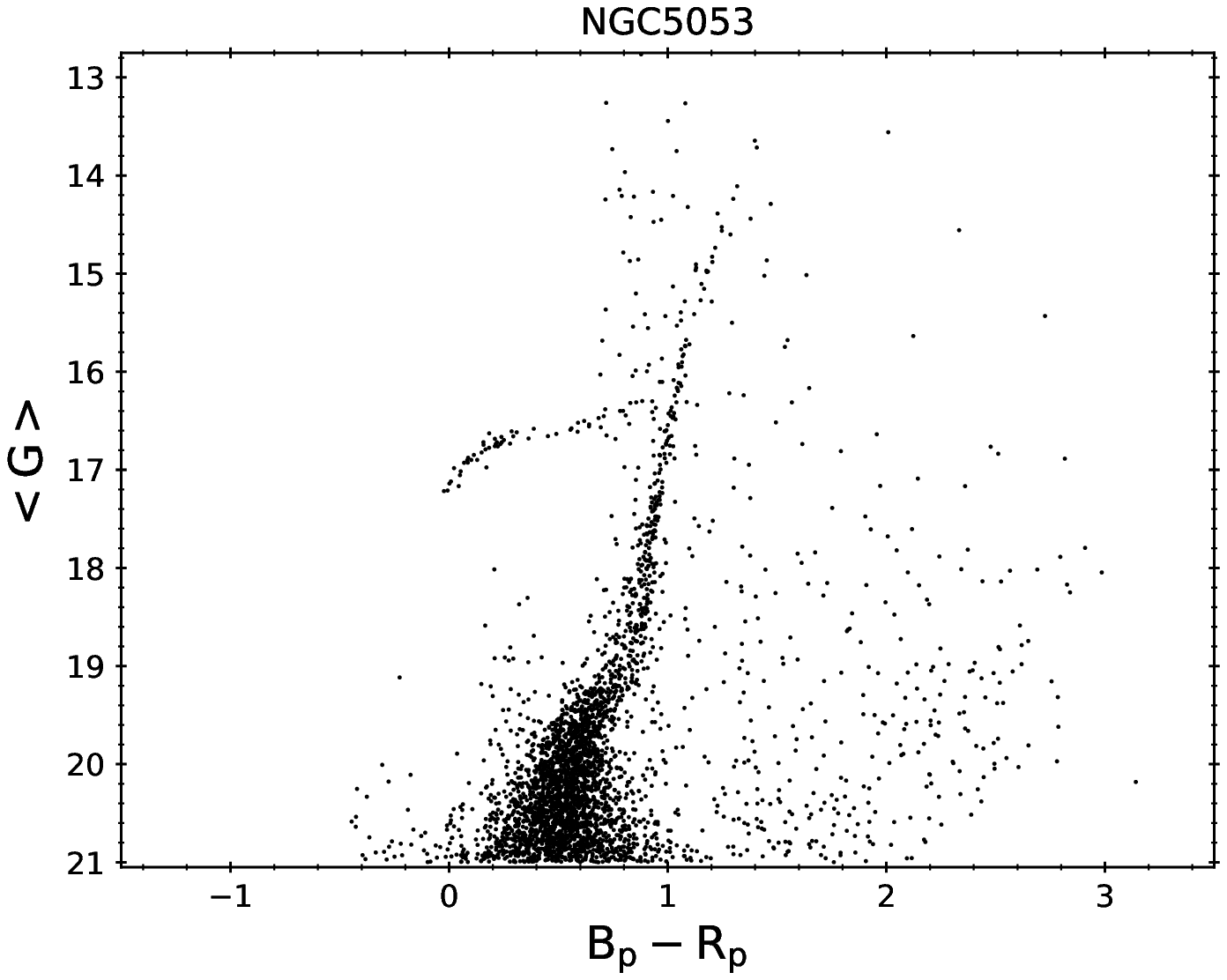}
  \caption{Gaia sources, queried from the DR2 main catalog, located within 1 tidal radius of NGC~5024 and NGC~5053. The left panels present the proper motions of these sources, while the right panels are the corresponding CMD. The dashed lines in left panels are the measured proper motions for each clusters adopted from \citet{gaia2018gc}, while the dashed circles represent the selection criterion of $\Delta_{\mathrm{PM}} = \mathrm{pmT}/2$ (only sources within the circles were be used to construct the ``clean'' CMD).}\label{fig_initial}
\end{figure*}

We utilize the proper motions and color-magnitude diagrams (PMD and CMD, respectively) to evaluate the probable membership of the selected RR Lyrae and candidates (as described in Section 2) in the vicinity of the two globular clusters studied in this work. More precisely, these diagrams, constructed from using the data in Gaia DR2 catalog, are used to reject RR Lyrae and candidates that are obviously not members of the clusters. Sources within 1 tidal radius of each globular clusters were queried from the Gaia DR2 main catalog \citep{gaia2018}, the corresponding PMD and CMD were shown in Figure \ref{fig_initial}. As can be seen from the figure, there are large scatters in the PMD and many ``outliers'' in the CMD, which are probably foreground or background sources. Since our goal is not to establish the cluster membership of each Gaia sources, but construct a ``clean'' CMD to be compared with the RR Lyrae and candidates, we adopted a simple criterion to select the Gaia sources. For each Gaia sources, we calculated $\Delta_{\mathrm{PM}} = [(\mathrm{pmRA}-\mathrm{pmRA}_c)^2 + (\mathrm{pmDE}-\mathrm{pmDE}_c]^2)^{1/2}$, where $\mathrm{pmRA}$ and $\mathrm{pmDE}$ are proper motions for each sources in right ascension and declination, respectively, and the subscript $_c$ represent the measured proper motions of the clusters \citep{gaia2018gc}. Only those sources with $\Delta_{\mathrm{PM}} < \mathrm{pmT}/2$ are kept as the ``clean'' sample, where $\mathrm{pmT}$ is the quadrature sum of $\mathrm{pmRA}_c$ and $\mathrm{pmDE}_c$. The ``clean'' CMDs of NGC~5024 and NGC~5053 are presented in Section 2. 

\section{B. Light Curves for Two Mis-Identified RR Lyrae}

\begin{figure*}
  \epsscale{1.2}
  \plotone{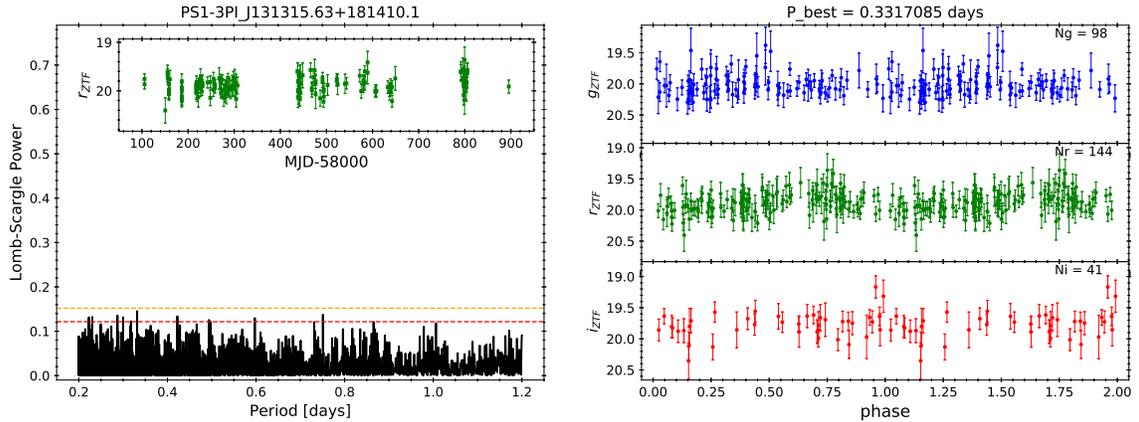}
  \caption{{\it Left panel:} The Lomb-Scargle periodogram for PS1-3PI J131315 based on the $r$-band ZTF light curve, shown in the inset figure. The horizontal red and orange dashed lines represent the false alarm probability of 0.1 and 0.01, respectively. {\it Right panel:} Folded ZTF $gri$-band light curves with the ``best'' period, corresponding to the highest peak in the Lomb-Scargle periodogram. Note that the ``best'' period does not necessary represent the true period. }\label{fig_badLC_ps1}
\end{figure*}

\begin{figure*}
  \epsscale{1.2}
  \plotone{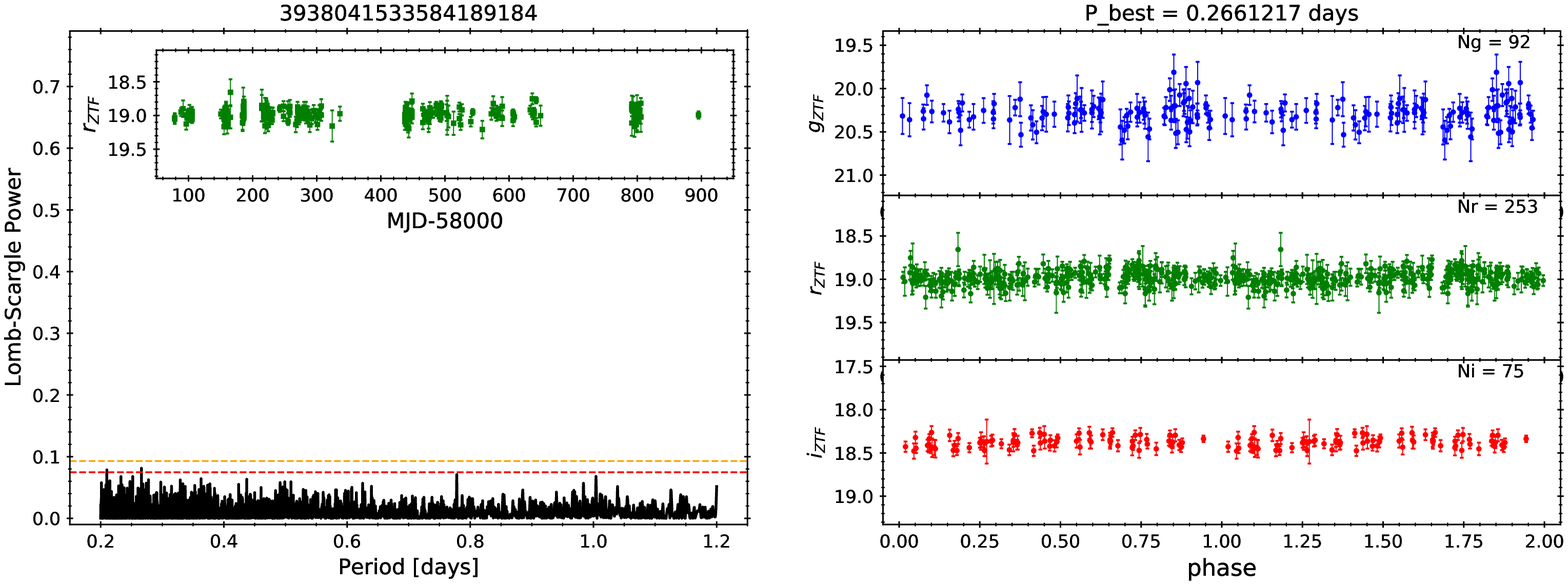}
  \caption{Same as Figure \ref{fig_badLC_ps1}, but for star with Gaia DR2 ID 3938041533584189184.}\label{fig_badLC_gaia}
\end{figure*}

We have examined the ZTF light curves for RR Lyrae listed in Table
\ref{tab2}, and found that they do not display RR Lyrae-like light
curves. These two putative RR Lyrae are PS1-3PI J131315.63+181410.1
(abbreviated as PS1-3PI J131315) and Gaia DR2 ID 3938041533584189184,
but we classify them as non-RR Lyrae. Other RR Lyrae in Table
\ref{tab2} display the light curve shapes expected for ab- or c-type RR Lyrae after folding the ZTF light curves with their published periods.
\begin{description}

\item[PS1-3PI J131315] This star is identified in \citet{sesar2017}
  with final classification scores of {\tt S3ab=0.93} and {\tt
    S3c=0.02}, suggesting this star has a high probability of being an
  ab-type RR Lyrae. The period of PS1-3PI J131315 is found to be
  0.6386368~days \citep{sesar2017}. However the Lomb-Scargle
  periodogram applied to its $r$-band ZTF light curve did not reveal
  any significant peak between 0.2 to 1.2 days (see left panel of
  Figure \ref{fig_badLC_ps1}). The folded ZTF light curves for this
  star, either with the published period or the ``best'' period
  (corresponding to the highest peak in the Lomb-Scargle periodogram),
  as shown in right panel of Figure \ref{fig_badLC_ps1}, do not
  resemble those of ab-type RR Lyrae.

\item[3938041533584189184] This star is identified in \citet{rimoldini2019} as an ab-type RR Lyrae with a {\tt best\_class\_score=0.6279}, but there is no period found for this star. We ran the Lomb-Scargle periodogram on the ZTF $r$-band light curve for this star, and no significant peak was found between 0.2 and 1.2~days. The Lomb-Scargle periodogram and the folded ZTF light curves with the ``best'' period are displayed in Figure \ref{fig_badLC_gaia}. Clearly, this star is not an ab-type RR Lyrae.
\end{description}

% ===============================================
%               REFERENCE
% ===============================================


\begin{thebibliography}{}

\bibitem[Arellano Ferro et al.(2011)]{af2011} Arellano Ferro, A., Figuera Jaimes, R., Giridhar, S., et al.\ 2011, \mnras, 416, 2265
    
\bibitem[Astropy Collaboration et al.(2013)]{astropy2013} Astropy Collaboration, Robitaille, T.~P., Tollerud, E.~J., et al.\ 2013, \aap, 558, A33

\bibitem[Astropy Collaboration et al.(2018)]{astropy2018} Astropy Collaboration, Price-Whelan, A.~M., Sip{\H{o}}cz, B.~M., et al.\ 2018, \aj, 156, 123

\bibitem[Baumgardt \& Makino(2003)]{baumgardt2003} Baumgardt, H., \& Makino, J.\ 2003, \mnras, 340, 227

\bibitem[Beccari et al.(2008)]{beccari2008} Beccari, G., Lanzoni, B., Ferraro, F.~R., et al.\ 2008, \apj, 679, 712
  
\bibitem[Bellm et al.(2019)]{bel19} Bellm, E.~C., Kulkarni, S.~R., Graham, M.~J., et al.\ 2019, \pasp, 131, 018002

\bibitem[Belokurov et al.(2006)]{belokurov2006} Belokurov, V., Evans, N.~W., Irwin, M.~J., et al.\ 2006, \apjl, 637, L29

\bibitem[Capuzzo Dolcetta et al.(2005)]{cd2005} Capuzzo Dolcetta, R., Di Matteo, P., \& Miocchi, P.\ 2005, \aj, 129, 1906
    
\bibitem[Chun et al.(2010)]{chun2010} Chun, S.-H., Kim, J.-W., Sohn, S.~T., et al.\ 2010, \aj, 139, 606
  
\bibitem[Clement et al.(2001)]{clement2001} Clement, C.~M., Muzzin, A., Dufton, Q., et al.\ 2001, \aj, 122, 2587

\bibitem[Clement(2017)]{clement2017} Clement, C.~M.\ 2017, VizieR Online Data Catalog, V/150

\bibitem[Clementini et al.(2019)]{clementini2019} Clementini, G., Ripepi, V., Molinaro, R., et al.\ 2019, \aap, 622, A60

\bibitem[Combes et al.(1999)]{combes1999} Combes, F., Leon, S., \& Meylan, G.\ 1999, \aap, 352, 149

\bibitem[de Boer et al.(2019)]{deboer2019} de Boer, T.~J.~L., Gieles, M., Balbinot, E., et al.\ 2019, \mnras, 485, 4906

\bibitem[Dehnen et al.(2004)]{dehnen2004} Dehnen, W., Odenkirchen, M., Grebel, E.~K., et al.\ 2004, \aj, 127, 2753
  
\bibitem[Dekany et al.(2020)]{dec20} Dekany, R., Smith, R.~M., Riddle, R., et al.\ 2020, \pasp, 132, 038001

\bibitem[Fellhauer et al.(2007)]{fellhauer2007} Fellhauer, M., Evans, N.~W., Belokurov, V., et al.\ 2007, \mnras, 380, 749
  
\bibitem[Fern{\'a}ndez-Trincado et al.(2015)]{ft2015} Fern{\'a}ndez-Trincado, J.~G., Vivas, A.~K., Mateu, C.~E., et al.\ 2015, \aap, 574, A15

\bibitem[Gaia Collaboration et al.(2016)]{gaia2016} Gaia Collaboration, Prusti, T., de Bruijne, J.~H.~J., et al.\ 2016, \aap, 595, A1
  
\bibitem[Gaia Collaboration et al.(2018a)]{gaia2018} Gaia Collaboration, Brown, A.~G.~A., Vallenari, A., et al.\ 2018a, \aap, 616, A1

\bibitem[Gaia Collaboration et al.(2018b)]{gaia2018gc} Gaia Collaboration, Helmi, A., van Leeuwen, F., et al.\ 2018b, \aap, 616, A12

\bibitem[Ginsburg et al.(2019)]{astroquery2019} Ginsburg, A., Sip{\H{o}}cz, B.~M., Brasseur, C.~E., et al.\ 2019, \aj, 157, 98
  
\bibitem[Graham et al.(2019)]{gra19} Graham, M.~J., Kulkarni, S.~R., Bellm, E.~C., et al.\ 2019, \pasp, 131, 078001

\bibitem[Grillmair(2019)]{grillmair2019} Grillmair, C.~J.\ 2019, \apj, 884, 174
  
\bibitem[Harris(1996)]{harris1996} Harris, W.~E.\ 1996, \aj, 112, 1487 

\bibitem[Harris(2010)]{harris2010} Harris, W.~E.\ 2010, arXiv:1012.3224 

\bibitem[Hozumi \& Burkert(2015)]{hozumi2015} Hozumi, S., \& Burkert, A.\ 2015, \mnras, 446, 3100

\bibitem[Ibata et al.(2017)]{ibata2017} Ibata, R.~A., Lewis, G.~F., Thomas, G., et al.\ 2017, \apj, 842, 120

\bibitem[Jordi \& Grebel(2010)]{jordi2010} Jordi, K., \& Grebel, E.~K.\ 2010, \aap, 522, A71

\bibitem[Kharchenko et al(2013)]{kharchenko2013} Kharchenko, N.~V., Piskunov, A.~E., Schilbach, E., et al.\ 2013, \aap, 558, A53

\bibitem[Kunder et al.(2018)]{kunder2018} Kunder, A., Mills, A., Edgecomb, J., et al.\ 2018, \aj, 155, 171
  
\bibitem[Kundu et al.(2019)]{kundu2019} Kundu, R., Minniti, D. \& Singh, H. P.\ 2019, \mnras, 483, 1737

\bibitem[Lauchner et al.(2006)]{lauchner2006} Lauchner, A., Powell, W.~L., \& Wilhelm, R.\ 2006, \apjl, 651, L33
  
\bibitem[Lee et al.(2006)]{lee2006} Lee, K.~H., Lee, H.~M., \& Sung, H.\ 2006, \mnras, 367, 646
  
\bibitem[Lehmann, \& Scholz(1997)]{lehmann1997} Lehmann, I., \& Scholz, R.-D.\ 1997, \aap, 320, 776

\bibitem[Masci et al.(2019)]{mas19} Masci, F.~J., Laher, R.~R., Rusholme, B., et al.\ 2019, \pasp, 131, 018003
  
\bibitem[McLaughlin \& van der Marel(2005)]{mvdm2005} McLaughlin, D. E. \& van der Marel, R. P.\ 2005, \apjs, 161, 304

\bibitem[Minniti et al.(2018)]{minniti2018} Minniti, D., Fern{\'a}ndez-Trincado, J.~G., Ripepi, V., et al.\ 2018, \apjl, 869, L10

\bibitem[Montuori et al.(2007)]{montouri2007} Montuori, M., Capuzzo-Dolcetta, R., Di Matteo, P., et al.\ 2007, \apj, 659, 1212

\bibitem[Myeong et al.(2017)]{myeong2017} Myeong, G.~C., Jerjen, H., Mackey, D., et al.\ 2017, \apjl, 840, L25

\bibitem[Navarrete et al.(2017)]{navarrete2017} Navarrete, C., Belokurov, V., \& Koposov, S.~E.\ 2017, \apjl, 841, L23

\bibitem[Nemec(2004)]{nemec2004} Nemec, J.~M.\ 2004, \aj, 127, 2185

\bibitem[Niederste-Ostholt et al.(2010)]{no2010} Niederste-Ostholt, M., Belokurov, V., Evans, N.~W., et al.\ 2010, \mnras, 408, L66
    
\bibitem[Odenkirchen et al.(2001)]{odenkirchen2001} Odenkirchen, M., Grebel, E.~K., Rockosi, C.~M., et al.\ 2001, \apjl, 548, L165
  
\bibitem[Peterson, \& King(1975)]{peterson1975} Peterson, C.~J., \& King, I.~R.\ 1975, \aj, 80, 427

\bibitem[Poleski(2014)]{poleski2014} Poleski, R.\ 2014, \pasp, 126, 509
  
\bibitem[Price-Whelan et al.(2019)]{pw2019} Price-Whelan, A.~M., Mateu, C., Iorio, G., et al.\ 2019, \aj, 158, 223

\bibitem[Rimoldini et al.(2019)]{rimoldini2019} Rimoldini, L., Holl, B., Audard, M., et al.\ 2019, \aap, 625, A97

\bibitem[Rockosi et al.(2002)]{rockosi2002} Rockosi, C.~M., Odenkirchen, M., Grebel, E.~K., et al.\ 2002, \aj, 124, 349

\bibitem[Sarajedini \& Milone(1995)]{sm1995} Sarajedini, A., \& Milone, A.~A.~E.\ 1995, \aj, 109, 269

\bibitem[Sesar et al.(2017)]{sesar2017} Sesar, B., Hernitschek, N., Mitrovi{\'c}, S., et al.\ 2017, \aj, 153, 204

\bibitem[Sollima et al.(2011)]{sollima2011} Sollima, A., Mart{\'\i}nez-Delgado, D., Valls-Gabaud, D., et al.\ 2011, \apj, 726, 47

\bibitem[Trager et al.(1995)]{trager1995} Trager, S.~C., King, I.~R., \& Djorgovski, S.\ 1995, \aj, 109, 218
  
\bibitem[Vivas et al.(2001)]{vivas2001} Vivas, A.~K., Zinn, R., Andrews, P., et al.\ 2001, \apjl, 554, L33

\bibitem[Vivas, \& Zinn(2006)]{vivas2006} Vivas, A.~K., \& Zinn, R.\ 2006, \aj, 132, 714

\bibitem[Watson et al.(2006)]{watson2006} Watson, C.~L., Henden, A.~A., \& Price, A.\ 2006, Society for Astronomical Sciences Annual Symposium, 25, 47
  
\bibitem[Wu et al.(2005)]{wu2005} Wu, C., Qiu, Y.~L., Deng, J.~S., et al.\ 2005, \aj, 130, 1640

\bibitem[Yang \& Sarajedini(2010)]{yang2010} Yang, S.-C., \& Sarajedini, A.\ 2010, \apj, 708, 293

\bibitem[Yim \& Lee(2002)]{yim2002} Yim, K.-J., \& Lee, H.~M.\ 2002, Journal of Korean Astronomical Society, 35, 75

\end{thebibliography}
\end{document}